**Systematically Dissecting the Global Mechanism of miRNA Functions in Pluripotent Stem Cells**

Anyou Wang[1*], et al

*Corresponding author
Anyou Wang
anyou.wang@alumni.ucr.edu

**Running title:** Global mechanisms of miRNA roles in stem cells

**Abstract:**

MicroRNAs (miRNAs) critically modulate stem cell properties like pluripotency, but the fundamental mechanism remains largely unknown. This study systematically analyzes multiple-omics data and builds a systems physical network including genome-wide interactions between miRNAs and their targets to reveal the systems mechanism of miRNA functions in mouse pluripotent stem cells. Globally, miRNAs directly repress the pluripotent core factors during differentiation state. Surprisingly, during pluripotent state, the top important miRNAs do not directly regulate the pluripotent core factors as thought, but they only directly target the pluripotent signal pathways and directly repress developmental processes. Furthermore, at pluripotent state miRNAs predominately repress DNA methyltransferases, the core enzymes for DNA methylation. The decreasing methylation repressed by miRNAs in turn activates the top miRNAs and pluripotent core factors, creating an active circuit system to modulate pluripotency. MiRNAs vary their functions with different stem cell states. While miRNAs directly repress pluripotent core factors to facilitate the differentiation during cell differentiation, they also help stem cells to maintain pluripotency by activating pluripotent cores through directly repressing DNA methylation systems and primarily inhibiting development.

## Introduction

MicroRNAs (miRNAs), short (~22 nts) conserved endogenous non-coding RNAs, inhibit messenger RNA targets by repressing translation or reducing mRNA stability [1] . MiRNAs critically modulate many cellular events, including the balance between proliferation and differentiation during organ development [1]. In pluripotent stem cells (including induced pluripotent stem cells and embryonic stem cells, referred to as stem cells hereafter), miRNAs play important roles in regulating stem cell bioprocesses [2-6].

miRNAs modulate stem cell pluripotency and differentiation [2-4]. Knocking out the key miRNA processing enzymes Dicer [2-3] or DGCR8 [4] causes stem cells to lose their pluripotency. MiRNA-290 cluster has been proposed to regulate the core pluripotency factors like POU5F1 [7-9]. MiRNA-302/367 cluster has also been used to induce pluripotency [10]. On the other hand, miRNAs like let-7 induce stem cell differentiation [11]. However, these recent studies have mostly focused on individual gene functions in stem cells although genome-wide data might be employed, and the conclusions drawn from these current studies are unavoidably biased on genes selected by these studies. Therefore, these studies only provide partial mechanisms of miRNA functions in stem cells, and the overall systems mechanisms of how miRNAs regulate stem cell processes remain largely elusive.

MiRNAs generally do not work alone to perform their functions [12]. One miRNA might target more than 100 genes [13-14], and one gene can be repressed by multiple miRNAs in a sequence-specific fashion [12-13, 15]. In turn, proteins can physically bind to the promoters and enhancers of miRNAs to regulate miRNA activations [16]. These binary interactions between miRNAs and proteins would form a complex systematic network. This complexity of miRNA interaction network make it challenged for conventional approaches like gene-knockout to unbiasedly capture the real mechanisms of miRNA functions in stem cells.

This present study employed systems physical network approaches [17] and constructed a comprehensive and unbiased map of genome-wide interactions between miRNAs and their targets to investigate the global basis of miRNA roles in pluripotent stem cells. Results of the present study lay a conceptual framework for future studies and applications of miRNAs in stem cells.

## Results

### Physical network of miRNA and protein interactions in stem cells

To systematically reveal the roles of miRNAs in stem cells, this study first constructed a systems network [17] of interactions between miRNAs and proteins. These interactions contain binary interactions from two directions, from miRNAs to mRNAs coding for proteins, and from proteins to miRNA promoters and enhancers. The miRNA-targets were inferred from CLIP-seq data, which simultaneously identify miRNAs-mRNA interactions by measuring miRNA-Argonaute complexes [18-19] (materials and methods). Protein-targets were inferred from ChIP-seq, which measures protein interactions with DNA [20] (Figure 1, Table S1 and materials and methods). The CLIP-seq and ChIP-seq provide data of physical binding interactions. The systems interaction network constructed here includes physical interactions of both miRNA-targets and protein-targets.

The entire network contains ~10,000 nodes and ~233,000 interactions (Figure 1C) and is accessible online (........). Both ChIP-seq and CLIP-seq measures genome-wide targets and thus this network provides a global map of miRNA targets in stem cells. For example, the genome-wide interactions between miRNAs and the pluripotent core factors (NANOG, POU5F1 and SOX2) could be extracted from this network (Figure 1D). Due to its natural interactions generated from experimental data, this network would provide accurate interactions between miRNA and their targets, and results generated from this network would be reliable.

**The primary role of miRNAs in stem cells**

To understand the primary role of all miRNAs activated (up- or down-regulated when compared with somatic cells) at pluripotent state in stem cells, we searched for the biological functions of the network activated by miRNAs in stem cells. To avoid the dataset biases, we included different datasets of miRNAs and genes coding for proteins and selected miRNAs and genes that are only activated with high frequency in all datasets (Table S2-S3, materials and methods). These activated miRNAs and genes were used to enrich the entire network (Figure 1) to get the activated network using methods as previously described [17]. The network activated by overexpressed miRNAs and down-regulated proteins (Table S2-S3) formed a subnetwork activated by up-regulated miRNAs in stem cells. We run the GO (gene ontology) functional analysis (www.**geneontology**.org/) of this subnetwork [21] by separately using references of both entire GO annotation database and a set of all up-regulated genes in stem cells to get the less biased enrichment results. Different references generated different enrichment p-value but produced the similar result that the entire subnetwork primarily functions for development, with corrected p-value<6.517e-62 (Bonferroni correction using entire annotation, referred as corr, hereafter) (Figure S1A). This activated subnetwork was further enriched by the developmental GO term to obtain the developmental module (Figure 2A). This developmental module was decomposed into functional modules based on network topology [22] and it contained 6 sub-functional modules (Materials and Methods). All of these 6 modules primarily function for development (corr p-value<8.2615E-83, Figure 2A), indicating that the primary function of over-expressed miRNAs at pluripotent state is to repress developmental modules.

The repressing role of up-regulated miRNAs was further evidenced by examining the targets of three primarily represented miRNA groups, the top overexpressed miRNA group, a miR-302/367 cluster, and a single miR-294. First, a total of 17 out of the 20 most important miRNAs, which were selected on the basis of the variance contribution to the system (material and methods, Table S4), directly target a developmental sub-network (corr p<6.3512E-20, Figure S2). Among the 17 miRNAs, the top 5 miRNAs also target a module that primarily functions for development (corr p<8.5158E-23, Figure 2B-2C). Furthermore, the well-known miR-302 cluster and even a single miRNA, miR-294, also target modules functionally enriched in the developmental category with respectively corr p-value <1.4436E-21 (Figure 3A-3B), and corr p-value<1.0157E-29 (Figure 3C-3D). Therefore, overexpressed miRNAs at pluripotent state primarily repress development. Biologically, to maintain the self-renewal and pluripotency, stem cells have some ways to prevent development and differentiation. This repressing function of miRNAs can help miRNAs claim their contributions to the stem cell properties at the pluripotent state.

On the other hand, the down-regulated miRNAs in stem cells directly target genes that primarily function for metabolism and pluripotency (corr-p<3.5159E-42, Figure S1B, Figure S3). These down-regulated miRNAs become up-regulated during differentiation and directly inhibit metabolism and pluripotency during this state.  Together, miRNAs primarily and directly repress development during the pluripotent state while they repress metabolism and pluripotency during cell differentiation.

**Paths from activated miRNAs to pluripotent core factors**

MiRNAs like miR-302 cluster mediate pluripotency [7, 23], and it thus was assumed that top over-expressed miRNAs such as miR-302 and miRNA-290-295 cluster in stem cells might directly or indirectly target pluripotent core factors [7, 23]. To investigate if these top miRNAs directly target the core factors, we systematically searched the shortest paths respectively from the top important miRNAs as described above (Materials and Methods, Table S4) to the three core factors (POU5F1, NANOG, and SOX2). Surprisingly, all these top miRNAs, including miR-302b, miR-367, miR-294, and miR-292, do not directly target any core factors (Figure 4A-4D). Actually, the direct basis of these miRNAs regulating the pluripotent core factors seemed blurred because all these miRNAs must go through at least 2 steps to reach any pluripotency core factor. These two steps include a miRNA and a protein, which are not consistently expressed with these miRNAs. This suggested that these top over-expressed miRNAs in stem cells do not directly mediate the pluripotency.

We then globally and unbiasedly searched for up-regulated miRNAs that target the pluripotent core factors (NANOG, POU5F1 and SOX2). Surprisingly, we only found one miRNA (miR-684) that barely up-regulated (~2 fold changed) in stem cells and directly binds to SOX2 (Figure 4E), which was also targeted by down-regulated miRNA-431. This indicated that activated miRNAs (>100miRNAs) do not primarily and directly target the pluripotent core factors. On the other hand, only limited miRNAs, regardless of expression, target POU5F1, while many miRNAs target SOX2 although they might not over-express in stem cells (Figure 1D), suggesting that the connection from miRNAs to the pluripotent core factors primarily go through SOX2, in contrast with the current thought that miRNAs should primarily target POU5F1 [10], a key factor for reprogramming induced pluripotent stem cells.

## Pluripotent network targeted by activated miRNAs

We next expanded the pluripotent gene list to all pluripotent genes uncovered by Hu et al [24]. We still focused on the direct miRNA target and searched the first neighbor of all overexpressed miRNAs (Figure S4A) and all down-regulated miRNAs (Figure S4B). Most of these targets are shared by up- and down-regulated miRNAs (Figure 5A-5B), indicating that the primary pluripotent genes in stem cells are carefully modulated by multiple up- and down-regulated miRNAs. The function of the entire shared network is primarily for extrinsic signal pathways associated with pluripotency (Figure 5B). For example, the highly connected nodes APC, RAD21 and EIF4G2 are involved in Wnt signaling and mitotic cell cycle pathways. Signaling pathways with similar functions were also found in the network targeted by over-expressed miRNAs only (Figure 5C) and in modules directly targeted by the represented miRNAs cluster in stem cells, such as miR-302/367 cluster (Figure 5D) and miR-294 (Figure 5E). This indicated that miRNAs in stem cells primarily function for modulating the balance of pluripotent signal pathways instead of directly targeting pluripotent core factors. This suggests that these regulations driven by miRNAs might go through multiple steps to the pluripotent core factors.

In contrast to the up-regulated miRNAs in stem cells, down-regulated miRNAs directly target the core pluripotency factors (Figure S5), suggesting that these miRNAs inhibit the core factors for pluripotency to facilitate differentiation when these down-regulated miRNAs become up-regulated during differentiation. To summarize, miRNAs do not directly target pluripotent core factors during pluripotent state but miRNAs directly target and repress these core factors during differentiation.

**MiRNAs abundantly target epigenetic system**

The above results indicated that the number of miRNA binding (degree) to their targets (nodes) was very limited. The target with the highest degree, EIF4G2, was only attacked by ~20 up-regulated miRNAs (Figure 5C). It was expected that a certain group of nodes should be targeted by more than that. This drove us to further search the network hubs (the important nodes) in the entire network. We systematically ranked the miRNA targets by degree (miRNA directly binding only) and obtained the top hubs. The top hubs mostly function for RNA processing, but surprisingly, DNMT3A, a DNA-methyltransferase for de novo DNA methylation, was among the top hubs. DNMT3A actually holds more than 160 miRNA binding sites in 3'-UTR region based only on 8bp seed mapping and it was ranked within the top 1% of the up-regulated miRNA targets (Figure 6A). DNMT3A was even ranked higher than EIF4G2, the highest ranked node in the pluripotent genes (Figure 6A, Figure 5B-5C), indicating that DNMT3A should be a top important node in the network directly regulated by miRNAs in stem cells. This also indicated that miRNAs predominately target DNA methylation system, rather than the pluripotent genes. We extracted the network of DNMT3A directly targeted by miRNAs and found that the top miRNA clusters (Table S4), such as miR-302/367 and miR-290-295 cluster, were among the miRNAs that target DNMT3A (Figure 6B). Similarly, Many well-known miRNA clusters (e.g. miR-290-295 and miR-302) in stem cells also target DNMT1 (Figure 6C), an enzyme predominately responsible for methylation in hemimethylated CpG islands. Many down-regulated miRNAs also target DNMT3A and DNMT1 (Figure S6), but their attacks would lead to differentiation instead of maintaining pluripotency in stem cells. These abundant overexpressed miRNAs that target the methylation system suggest that miRNAs predominately repress DNMTs in stem cells.

In addition, miRNAs directly and abundantly target a core histone modification complex (HDAC4-MEF2C-MEF2D, http://www.ncbi.nlm.nih.gov/gene/9759) (Figure 7), including MEF2C (myocyte enhancer factor 2C), which was targeted by the top over-expressed miRNA clusters including miR-290-295 and miR-302 cluster (Figure 7). Up-regulating MEF2C enhances stem cells differentiation [25], and down-regulated MEF2C should inhibit differentiation. The down-regulation of MEF2C targeted by the top over-expressed miRNA clusters suggests that miRNAs repress differentiation in stem cells. This is consistent with our discussion above on the miRNA repressing development and differentiation at the pluripotent state in stem cells (Figure 2-3). Together, miRNAs directly and abundantly target the epigenetic systems at the pluripotent state.

**DNA methylation mediates the miRNA activation in stem cells**

To search the mechanism controlling the miRNA activations, this study turned to the genome-wide sequencing of DNA methylation in stem cells and methylation-loss-stem cells [26] (Table S1, materials and methods). A total of 2000bp in each upstream and downstream of start sites of all activated miRNAs were examined. While the DNA methylation in the downstream of up-regulated miRNAs is not different from that of down-regulated miRNAs ($p>0.1899$), the down-regulated miRNAs hold significantly higher methylation upstream than up-regulated miRNAs ($p<3.685e-05$, Figure 8A). Surprisingly, the biggest difference locates in ~1000bp up-stream instead of immediate up-stream ($p<1.265e-06$, Figure 8B). Furthermore, these differences are overall negatively correlated to miRNA expressions with correlation coefficient of -0.35 and p-value < 0.05 (Figure 8C). This suggested that the difference in DNA methylation accounts for the miRNA activations. This parallels a recent observation showing that the loss of DNA methylation significantly increases miRNA expressions [29]. Therefore, miRNA activations and their network are mediated by DNA methylation in ~1000bp upstream regions.

**Discussion**

This study is the first, to our knowledge, to investigate the primary mechanism of miRNA functions in stem cells at systems level on the basis of a physical map constructed by direct interactions of miRNAs and proteins. All data employed in this study were collected from published biological experiments, and the core part of this study, the miRNA-target physical binding map, was inferred from the CLIP-seq (Figure 1). This gives our study several advantages. First, the interaction map should be more accurate than that predicted by pure computations, a motif prediction [27], which only produced 20% overlapped with the experimental data (data not shown). Secondly, the nature of map inferred here provides the direct physical binding linkages between miRNAs and their targets. This network(....) makes it possible for us to understand a precise mechanistic picture of miRNA interactions with their targets. In contrast, traditional functional studies like gene knockdown/knockout only provide the linking hint from a gene to phenotypes with unknown mechanisms. Finally, the map provides a path to appreciate the molecular mechanics at systems level, like the pattern and network module recognitions that are based on many components (genes in this case). Mistakes could be made during observations of individual components (genes) because these individual observations could vary with conditions, but the nature of modules and patterns are normally robust enough to buffer the individual variations and noises, and they would not be changed with conditions [28]. Therefore, the results and conclusions on the global roles of miRNAs on the basis of pattern and module recognition here should represent the real nature of miRNA functions in stem cells.

Current studies have demonstrated that miRNAs play critical roles in maintaining overall properties like pluripotency in stem cells [2-4]. However, the mechanisms still remain elusive. In this study, we systematically revealed that one of miRNA primary functions is to repress developmental modules during the pluripotent state while miRNAs directly target pluripotent core factors during differentiation state (Figure 2, Figure 3, and Figure S3). This suggests that miRNAs primarily repress development at pluripotent state to prevent stem cell differentiation and to keep stem cell pluripotency while another set of miRNAs degrade pluripotent core factors to facilitate differentiation during differentiation state. This is consistent with the recent observation that overexpressions of miRNAs induce pluripotency [10] and miRNAs also facilitate stem cells differentiation [11].

The linkage between miRNAs and pluripotency has been widely investigated [7-9], but whether the linkage is direct or indirect still remains to be investigated. It has been consistently observed that gene expressions of the top over-expressed miRNAs are positive correlative to that of pluripotent core factors. Recent evidences also show that miRNA-302/367 cluster could induce pluripotency [10]. The observations lead to a speculation that miRNAs might directly target the pluripotent core factors. Although miRNAs could have many functions in certain conditions, the primary functions of miRNAs are for degrading and inhibiting their targets. If the overexpressed miRNAs directly target the core factors, these miRNAs would likely repress the core factors as previously evidenced [29], leading to down-regulations of these core factors. A negative correlation between them should show up, but the fact is that positive correlation has been consistently observed. This suggests that the top miRNAs might not directly target and degrade pluripotent core factors. Here, we utilized the power of our system network to exhaustively search the direct linkages between miRNAs and the pluripotent core factors. Our results revealed that the top miRNAs (Table S4) such as miR-290 and miRNA-302 cluster do not directly target any core pluripotent factors during the pluripotent state (Figure 4). Most of top miRNAs only directly target the extrinsic signal pathways associated with pluripotency (Figure 5). A signal from a signal pathway normally travels through multiple steps to reach its targets. Therefore, the top miRNAs indirectly target the core factors and indirectly regulate the pluripotency.

In contrast to the indirect mechanism of miRNAs in targeting the genetic system, the top miRNAs, including miRNA-302/367 and miRNA-294 cluster, directly and abundantly target the core enzymes of DNA methylation system, DNMT3A and DNMT1 (Figure 6). Targeting these DNMTs by the top expressed miRNAs suggests inhibition of DNMTs, which highly express in somatic cells but lowly express in stem cells (Figure 6). This parallels the most recent observations showing that gene expressions of up-regulated miRNAs are negatively correlative with that of DNA methyltransferases (DNMTs) [30-31] and that microRNAs degrade DNMTs in stem cells [30].

Our result of miRNA directly repressing DNMTs can help to understand the positive relationship existing between up-regulated miRNAs and overexpressed core pluripotent factors in stem cells as discussed above. Over-expressed miRNAs directly target DNMTs, leading to DNA methylation reductions at genome-wide level, including the loci of the pluripotent core factors. This results in over-expression of these core factors responsible for pluripotency. This is consistent with the observation of less methylation in the loci of the core factors during induced pluripotent stem cell reprogramming, and it is also consistent with our recent finding that the demethylation level can be used as a variable for discriminating different stem cells [32]. Therefore, miRNAs primarily and directly target the epigenetic system that further activates pluripotent core factors in stem cells. This parallels the most current report that miRNAs degrade DNMTs in stem cells [30].

DNA methylation might regulate expressions of a certain group of genes in stem cells [28]. Our data further showed that DNA methylation globally mediates the miRNA activations in stem cells (Figure 8). These miRNA activations by DNA methylation in turn repress the DNA methylation (Figure 6). Less methylation activates the miRNAs and pluripotent core factors again.

Together, we proposed a system-wide circuit to describe a part of miRNA primary roles in modulating pluripotency in pluripotent stem cells (Figure 9). In this circuit, miRNAs directly repress development and directly repress the DNA methylation system, while miRNAs indirectly regulate pluripotency genes. This repression of DNA methylation activates both pluripotent factors and miRNAs. The activations of pluripotent factors and repression of development contribute to pluripotency in stem cells, while the activations of miRNAs further inhibit both DNA methylation and development. This create an active system-wide circuit in stem cells to maintain the pluripotent state (Figure 9).

We here pay more attentions on the primary functions of miRNAs in pluripotent state than that in differentiation state and we do not exclude other functions of miRNAs in stem cells. The results made here are based on the current limited data and these miRNA functions drawn here only account for as a part of miRNA roles in stem cells.

With data accumulating, more functions of miRNAs will certainly be explored. However, understanding the fundamental systems roles of miRNAs studied here would broadly direct the future functional studies of miRNAs in stem cells and would guide the successful development of stem-cell–based therapies for regenerative medicine.

## Materials and methods

### Data resources

This study analyzed data generated by high-throughput methods, including CLIP-seq, ChIP-seq, RNA-seq, microarray, and bisulfite sequencing DNA (Table S1). The data was downloaded from GEO database (www.ncbi.nlm.nih.gov/geo/) and the details are shown in table S1.

### Data bioinformatics analysis

To be consistent and comparable, all sequencing data were mapped to mouse genome (mm9). All fastq files (Table S1) from CLIP-seq, ChIP-seq and RNA-seq were aligned using BWA 0.6.2 with default parameters [33]. All PCR-duplicates were removed. Bisulfite sequencing DNA was aligned by Bismark 0.7.6 with tolerating one non-bisulfite mismatch per read (http://www.bioinformatics.babraham.ac.uk/projects/bismark/).

The clusters from CLIP-seq were generated by using GenomicRanges 1.12 under R 2.15 and were subjected to second noise quartile cutoff. The filtered clusters located in 3'UTRs and CDSs were used to search for miRNA bindings. The miRNA bindings were searched against the perfect match of miRNA seed sequences (6-8nt) using home-made python scripts. The seed sequences were extracted from miRNA sequences downloaded from miRBase 19 (http://microrna.sanger.ac.uk/). Peaks from ChIP-seq were called using SISSRs 1.4 (http://dir.nhlbi.nih.gov/papers/lmi/epigenomes/sissrs/) and the peaks were annotated using ChIPpeakAnno 2.8 under R-2.15. The differential expressions from RNA-seq were performed using negative binomial model using edgeR 3.2 under R 2.15. Bisulfite DNA methylation was counted against the genome coordinates of miRNAs (Table S4). Microarray data were analyzed using limma 3.16 under R 2.15.

**Activated genes and miRNAs selection**

To minimize the biases from individual experiments and cell lines as well as the noises caused by high-throughput methods, we selected genes and miRNAs activated in different conditions by using different sets of data(Table S1). Expression data from various resources (Table S1) were employed. Genes coding for proteins with activating frequency >50% in all observations and miRNAs with activating frequency of more than 25% out of all observations were treated as activated genes and were selected (Table S2-S3). The up-regulation and down-regulation was based on comparison to somatic cells in each experiment (corrected $p< 0.01$ and fold change > 2).

The top important miRNAs (Table S4) were selected on the basis of their contributions to network structure and variance by using the algorithm as we previously published [28]. Briefly, the top miRNAs were selected on the basis of their ranking scores calculated by the eigengene-based connectivity as defined below [28].

$$SCORE = \frac{d_i}{d_{max}} + 2 \times cor|Xi, E)|$$

where $d_i$ denotes the $i^{th}$ node degree, and $d_{max}$ denotes the maximum degree of a node in the entire network. |Cor(xi, E)| represents the absolute Pearson correlation coefficient, where $x_i$ represents a vector of $i^{th}$ node value, and E eigengene of the network.

**Network construction and analysis**

The network construction and analysis were performed by approaches as our previous reports [17, 34]. Briefly, we built the map (Figure 1, ……..) by collecting the interactions of both miRNAs targets from the CLIP-seq and protein-binding promoters of miRNAs from ChIP-seq data. The interactions from CLIP-seq and ChIP-seq were signed as miRNA or proteins respectively (Figure 1). Only the direct interactions (first-neighbor) were selected and included, and thus this map is a physical binding network. The network was enriched by the activated genes and miRNAs selected above to get the activated network. Functional modules were further enriched by the functional genes based on gene ontology enrichment (e.g. Figure S1, http://www.geneontology.org/). Six sub-modules in functional developmental module (Figure 2A) were based on network topology to identify the densely connected modules. The target node ranking was based on the degree of each miRNA node (Figure 6A-6B).

## Acknowledgments

We specially thank Drs. Ying Du and Chunxiao Zhou for data collection.

## Author contributions

A.W, conceived and designed the experiments, analyzed the data, contributed data/materials/analysis tools, wrote the manuscript. Y.Z., Q.H. contributed data/materials/analysis tools. All authors reviewed the manuscript.

## Competing interests

No competing interests exist in this study

No financial conflict of interest in this study.

## Figure legends

**Figure 1. Overall view.** A, The concept of miRNA and protein interactions. B, Workflow of this study. C, Overall view of the entire network constructed by this study. D, A sample of entire network contents shows direct interactions between miRNAs and pluripotent core factors (NANOG, POU1F5, and SOX2). Nodes denote miRNAs or proteins as labeled; red node represents the gene up-regulation in pluripotent stem cells, green node as down-regulation, and white node as insignificant expression. Edges represent interactions; red edge represents miRNA targeting proteins and green edge represents the binding of proteins with the regulatory elements of miRNAs. This labeling strategy applies to all figures in this study.

**Figure 2. MiRNAs primarily repress developmental processes in pluripotent stem cells.** A, Up-regulated miRNAs primarily target a developmental module, which includes 6 primary sub-modules functioning for development as highlighted in cycles. B, The key modules targeted by the top 5 important miRNAs (Table S4). C, the key modules were significantly and abundantly enriched in the developmental category.

**Figure 3. Modules targeted by miRNA-302/367 cluster and miRNA-294.** A, modules targeted by the miRNA-302/367 cluster and their functional enrichment was shown in B. C, a module targeted by miRNA-294, and D, its functional enrichment.

**Figure 4. Pathways from activated miRNAs to pluripotent core factors.** A, B, C, D,

the shortest paths from miR-302b, miR-367, miR-294, and miR-292-5p respectively to the pluripotent core factors. None of these top overexpressed miRNAs directly binds to any of these core factors. E, Core factors directly targeted by activated miRNAs in stem cells.

**Figure 5. Pluripotent genes targeted by over-expressed miRNAs in stem cells.** A. Venn diagram of pluripotent genes targeted by both up-and down-regulated miRNAs. B, activated modules commonly targeted by both up- and down-regulated miRNAs. The enlarged nodes represent the highly connected targets that are targeted by both up- and down-regulated miRNAs in stem cells, and they work for pluripotent signal pathways. C, down-regulated modules targeted by over-expressed miRNAs in stem cells. D, activated modules targeted by the miRNA-302/367 cluster in stem cells. E, an activated network targeted by miR-294.

**Figure 6. MiRNAs abundantly target DNA methylation systems.** A, DNMT3A was targeted by 30 up-regulated miRNAs and was ranked in the top 1% of the up-regulated miRNA targets. The network nodes (miRNA targets) were plotted against the node degree (miRNA binding number only). B, the DNMT3A network targeted by miR NAs. The most important miRNAs, including the miRNA-302/367 and miR-290-295 cluster, are found in the DNMT3A network. C, miRNAs target another methylation enzyme DNMT1.

**Figure 7. MiRNAs directly target a histone complex.** The MiR-290-295 cluster heavily attacks the MEF2C.

**Figure 8. DNA methylation mediates miRNA activations in stem cells.** A, The methylation levels upstream and downstream from the miRNA start site. Down-regulated miRNAs have significantly higher methylation in upstream region when compared with up-regulated miRNAs. B, Detailed methylation profiling for regions that are 2000bp upstream from the activated miRNAs. The top 30 down-regulated miRNAs (Table S2) have higher methylation around -1000bp (green highlighted in middle panel) than that of the top 30 up-regulated miRNAs (Table S2, upper panel). The methylation profiling of a single down-regulated miRNA-133 as a representative example (bottom panel). C. Negative correlation between DNA methylation and miRNA expression. Red line represents regression line.

**Figure 9. A system view of miRNAs primary mechanistic roles in maintaining pluripotency in pluripotent stem cells.** MiRNAs directly repress both the DNA methylation system and the development, while indirectly regulate pluripotency genes. Reduced DNA methylation activates the miRNAs and pluripotent core factors for pluripotency. The activated circuit between miRNAs and DNA methylation, as well as the development inhibition, help stem cells to maintain the pluripotent state, see text for detail. The solid lines are directly derived from the present study and they represent direct interactions and the dash dark line represents indirect interactions, while dash green lines denote evidences from reference papers.

**Systematically Dissecting the Global Mechanism of MiRNA functions in Stem Cells**

Anyou Wang[1*], el at

**Supplemental materials**

Figure S1, Functional enrichments of all targets that are targeted by all up- and down-regulated miRNAs. A, Up-regulated miRNAs. B, down-regulated miRNAs.

Figure S2, total 17 out of top 20 important miRNAs target developmental genes.

Figure S3, down-regulated miRNAs mediate metabolism in stem cells.

Figure S4, all pluripotent genes targeted by up-regulated miRNAs (A) and down-regulated miRNAs (B) in stem cells.

Figure S5. An activated network targeted by down-regulated miRNAs.

Figure S6. Activated miRNAs targeting DNA methylation systems. A. Up-regulated miRNAs targeting DNMT3A. B, Down-regulated miRNAs targeting DNMT3A.

Table S1, Data sources

Table S2, a list of miRNAs that are frequently and significantly differential expressed between stem cells and somatic cells.

Table S3, a list of genes coding for proteins that are frequently and significantly differential expressed between stem cells and somatic cells.

Table S4, a list of top 20 important miRNAs

Figure 1

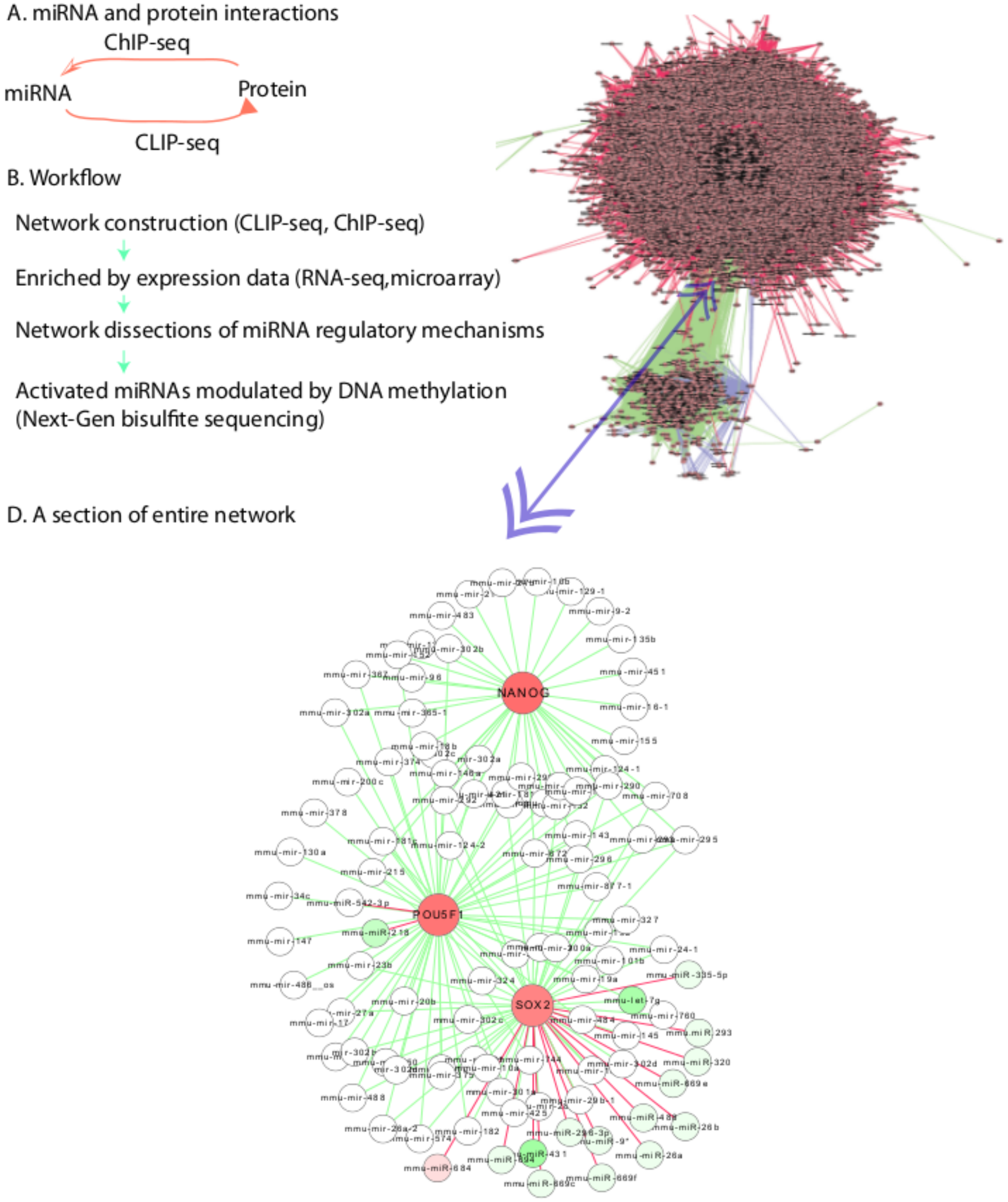
A. miRNA and protein interactions
ChIP-seq
miRNA
Protein
CLIP-seq
B. Workflow
Network construction (CLIP-seq, ChIP-seq)
Enriched by expression data (RNA-seq,microarray)
Network dissections of miRNA regulatory mechanisms
Activated miRNAs modulated by DNA methylation
(Next-Gen bisulfite sequencing)
C. Entire network
D. A section of entire network
NANOG
POU5F1
SOX2

Figure 2

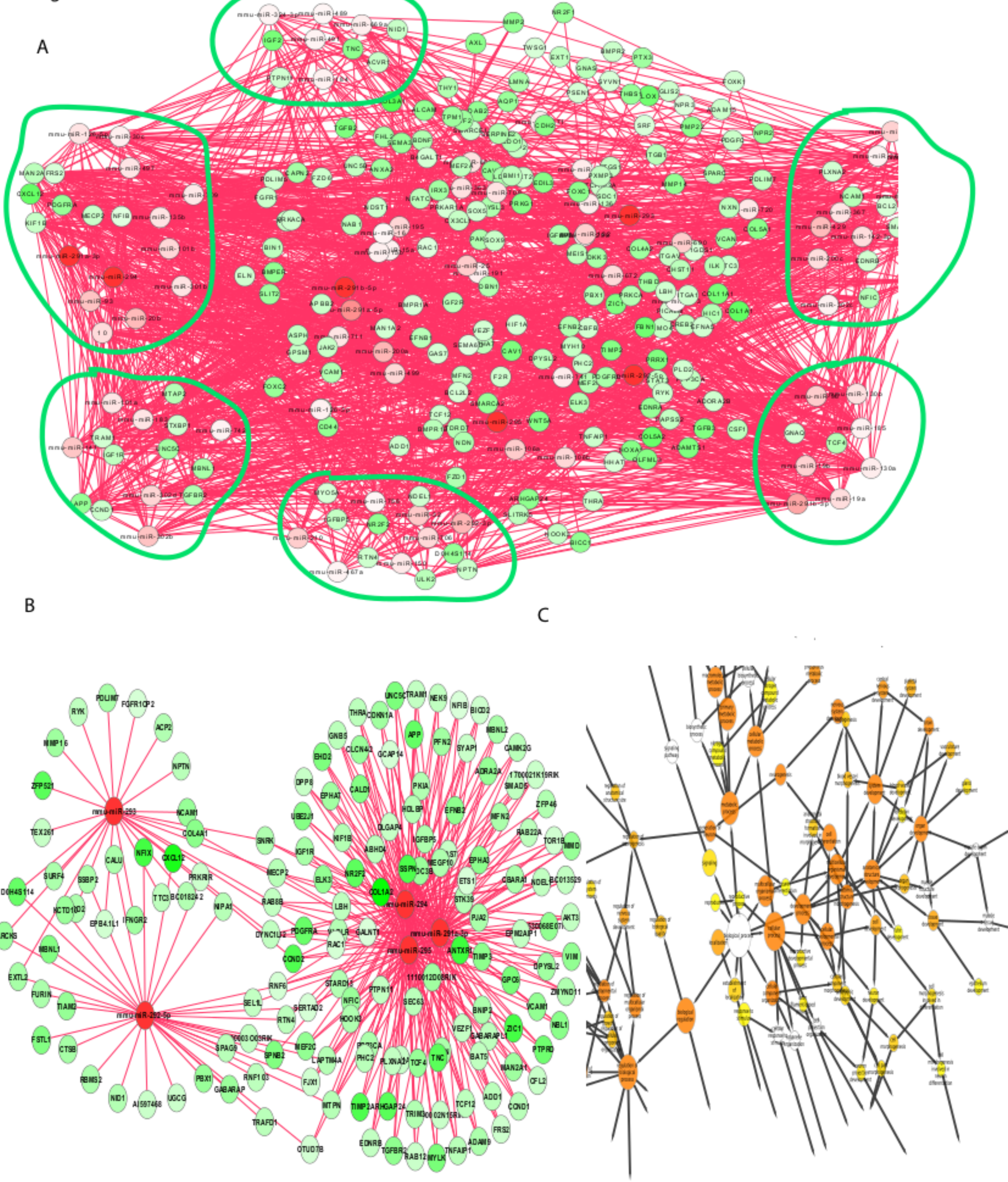

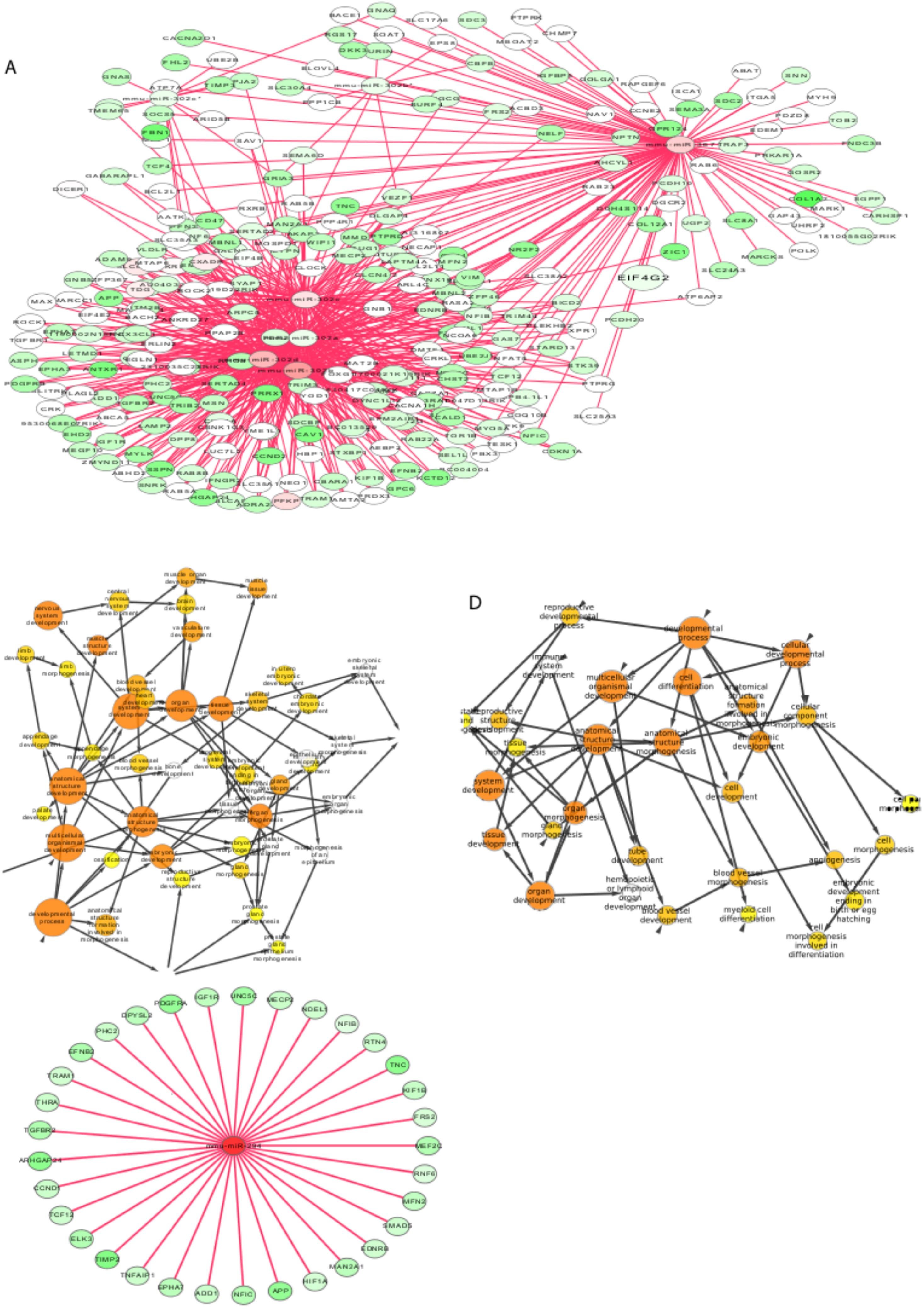
A
B
C
D
mmu-miR-302a
mmu-miR-302d
mmu-miR-302b*
mmu-miR-302c*
EIF4G2
nervous system development
muscle organ development
muscle tissue development
brain development
vasculature development
organ development
tissue development
developmental process
anatomical structure development
anatomical structure morphogenesis
multicellular organismal development
embryonic development
system development
ossification
reproductive developmental process
developmental process
cellular developmental process
immune system development
multicellular organismal development
cell differentiation
anatomical structure formation involved in morphogenesis
cellular component morphogenesis
reproductive structure development
tissue morphogenesis
anatomical structure development
anatomical structure morphogenesis
embryonic development
system development
organ morphogenesis
gland morphogenesis
cell development
tissue development
tube development
angiogenesis
cell morphogenesis
organ development
hemopoietic or lymphoid organ development
blood vessel development
blood vessel morphogenesis
myeloid cell differentiation
embryonic development ending in birth or egg hatching
cell morphogenesis involved in differentiation
mmu-miR-294
PDGFRA
IGF1R
UNC5C
MECP2
NDEL1
NFIB
RTN4
TNC
KIF1B
FRS2
MEF2C
RNF6
MFN2
SMAD5
EDNRB
MAN2A1
HIF1A
APP
NFIC
ADD1
EPHA7
TNFAIP1
TIMP2
ELK3
TCF12
CCND1
ARHGAP24
TGFBR2
THRA
TRAM1
EFNB2
PHC2
DPYSL2

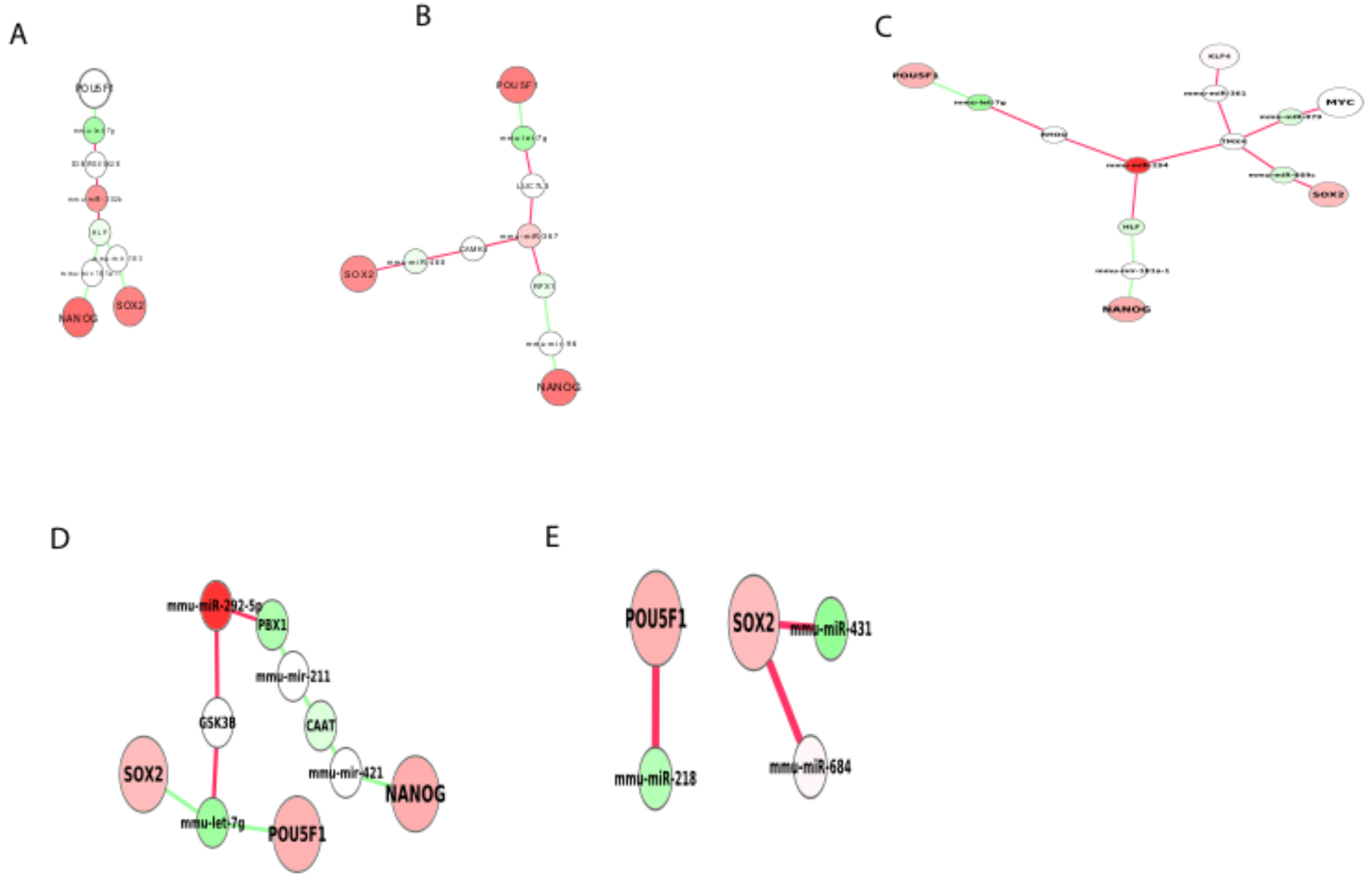

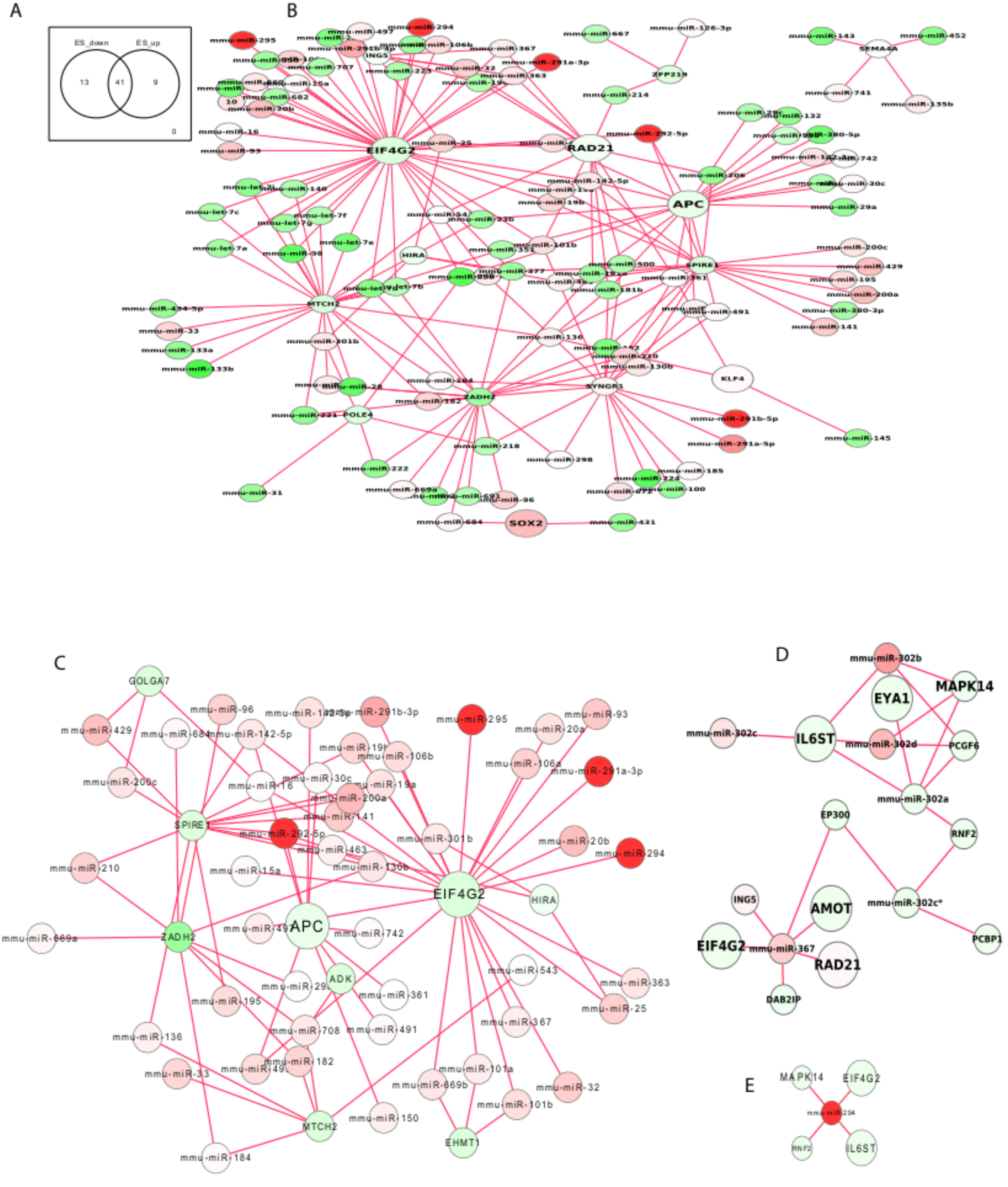

Figure 6

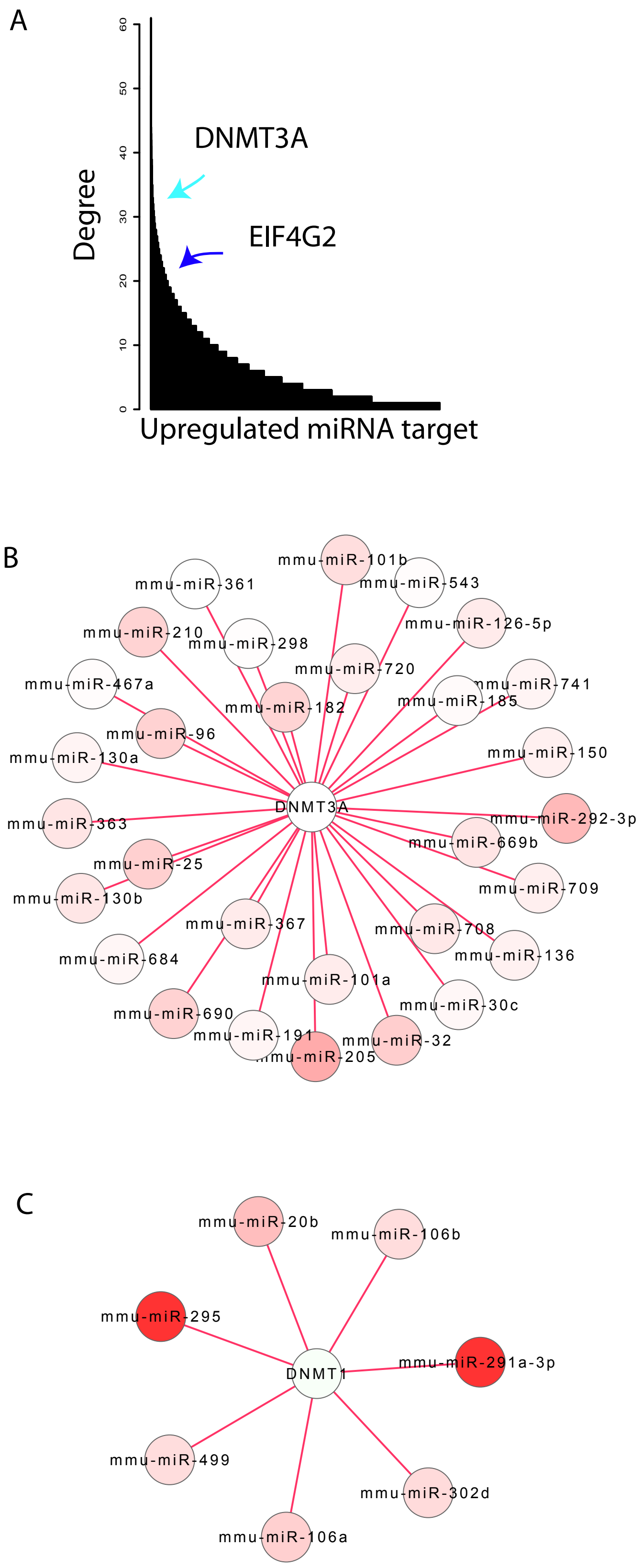

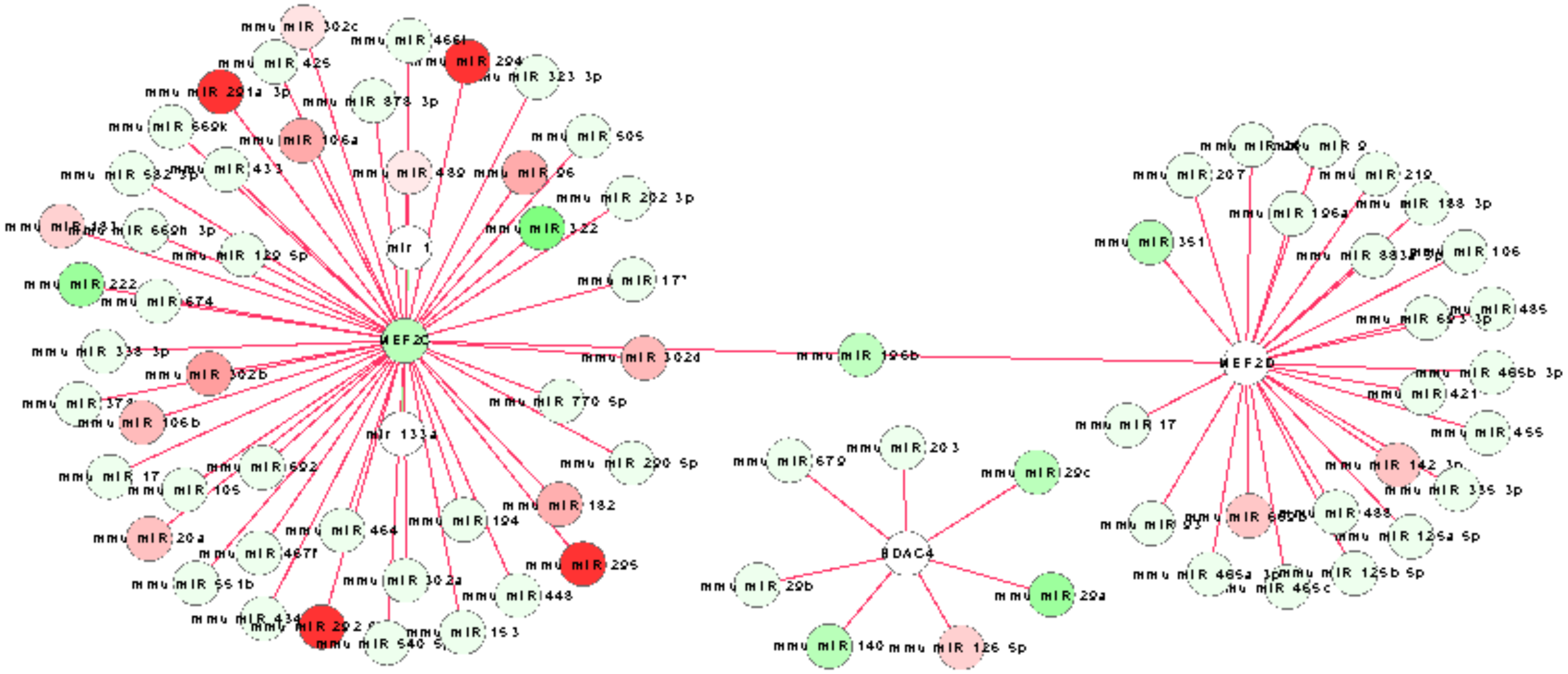

Figure 8

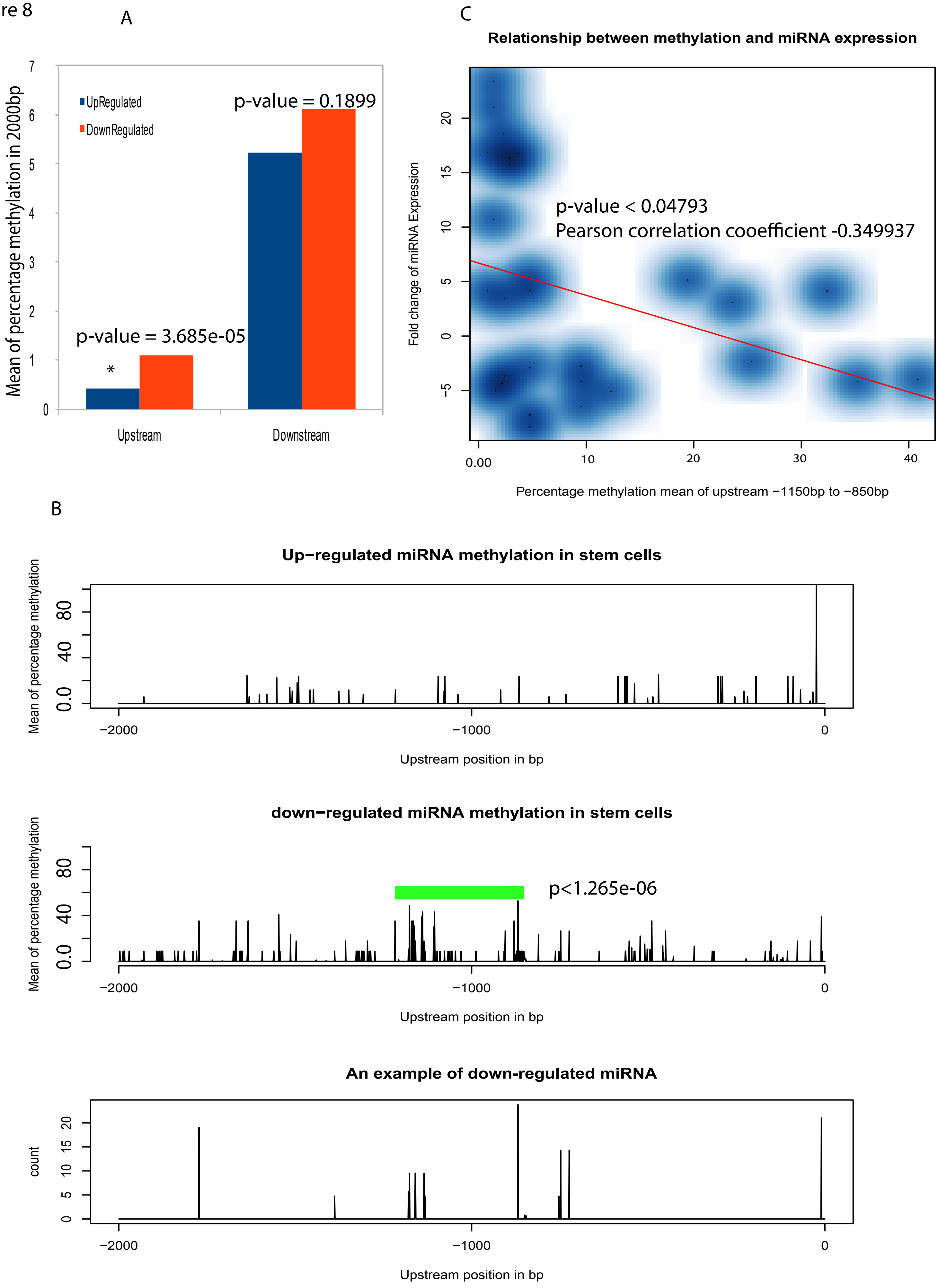
A
Mean of percentage methylation in 2000bp
UpRegulated
DownRegulated
p-value = 0.1899
p-value = 3.685e-05
*
Upstream
Downstream
C
Relationship between methylation and miRNA expression
Fold change of miRNA Expression
p-value < 0.04793
Pearson correlation cooefficient -0.349937
Percentage methylation mean of upstream −1150bp to −850bp
B
Up−regulated miRNA methylation in stem cells
Mean of percentage methylation
Upstream position in bp
down−regulated miRNA methylation in stem cells
p<1.265e-06
An example of down-regulated miRNA
count

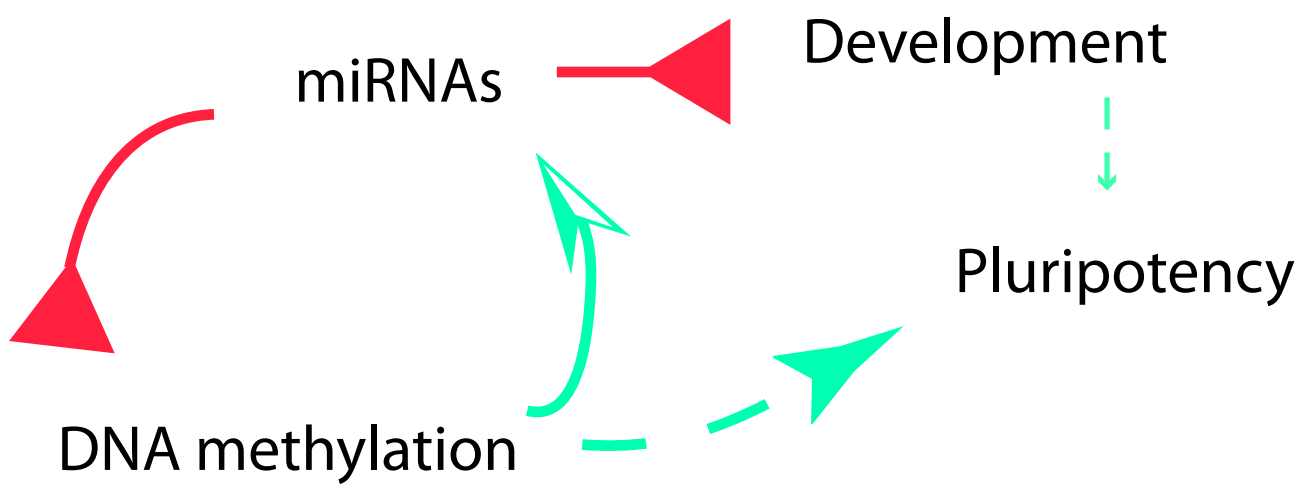

miRNAs
Development
Pluripotency
DNA methylation

A

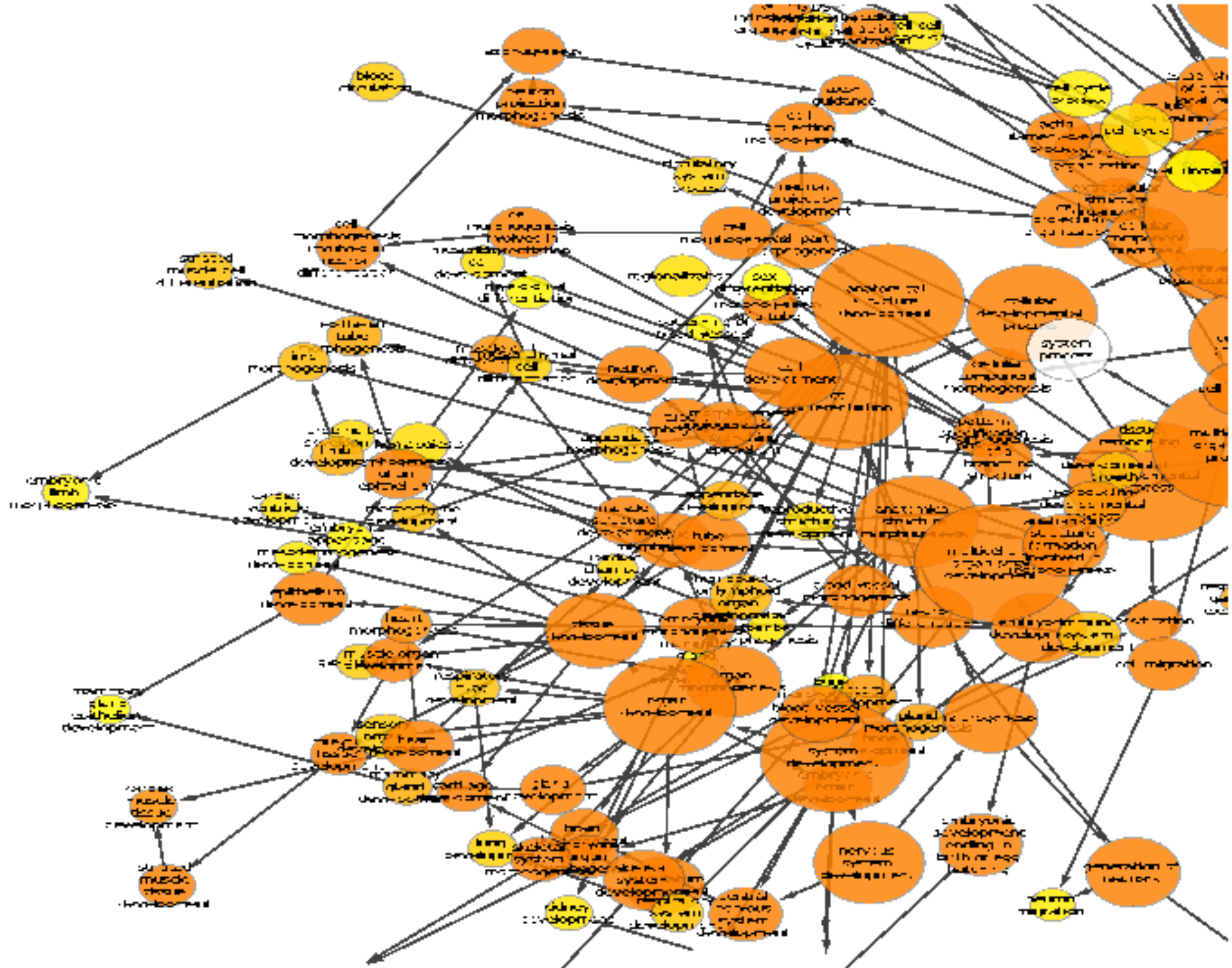

B

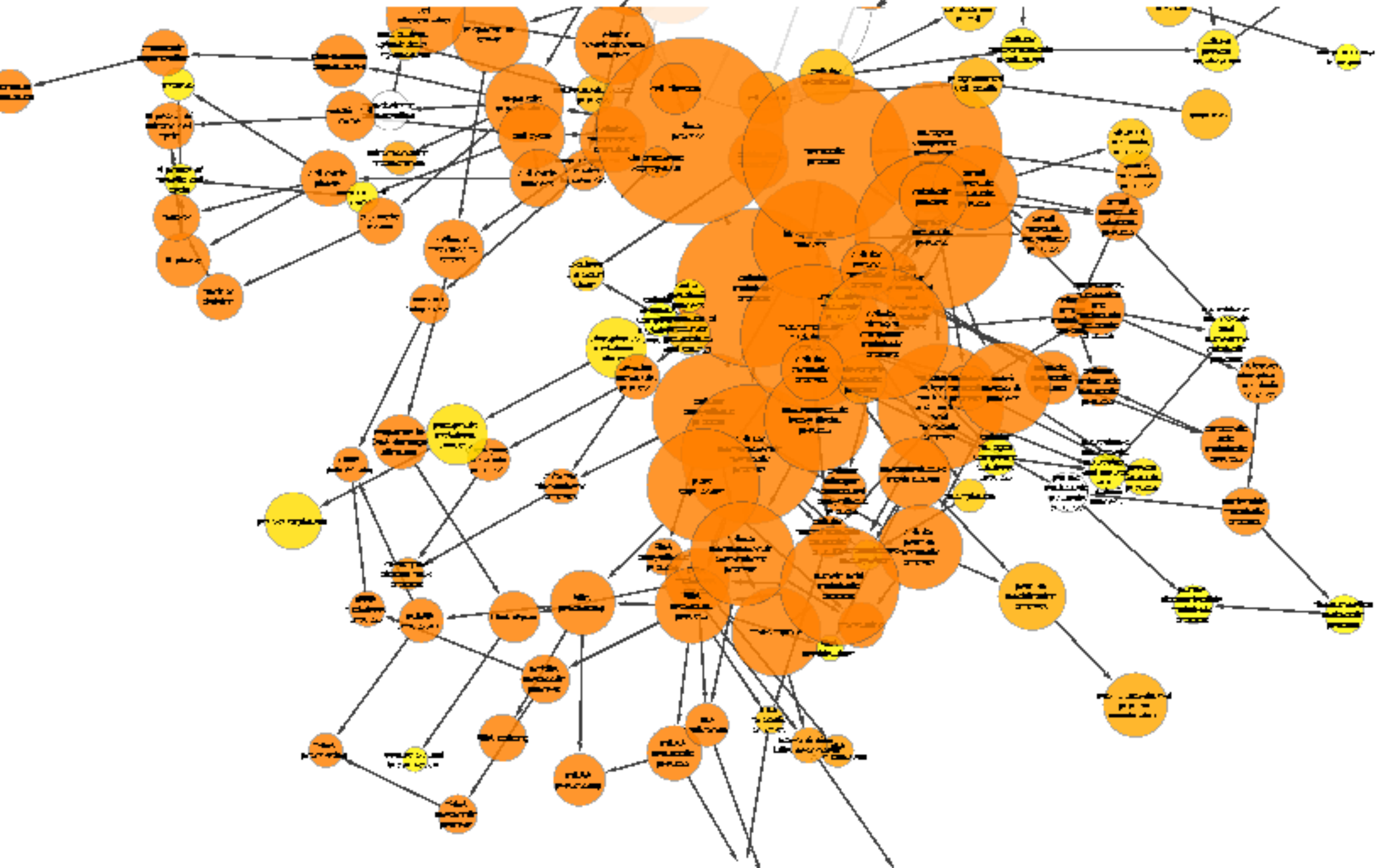

Total 17 out of top 20 miRNAs target developmental genes

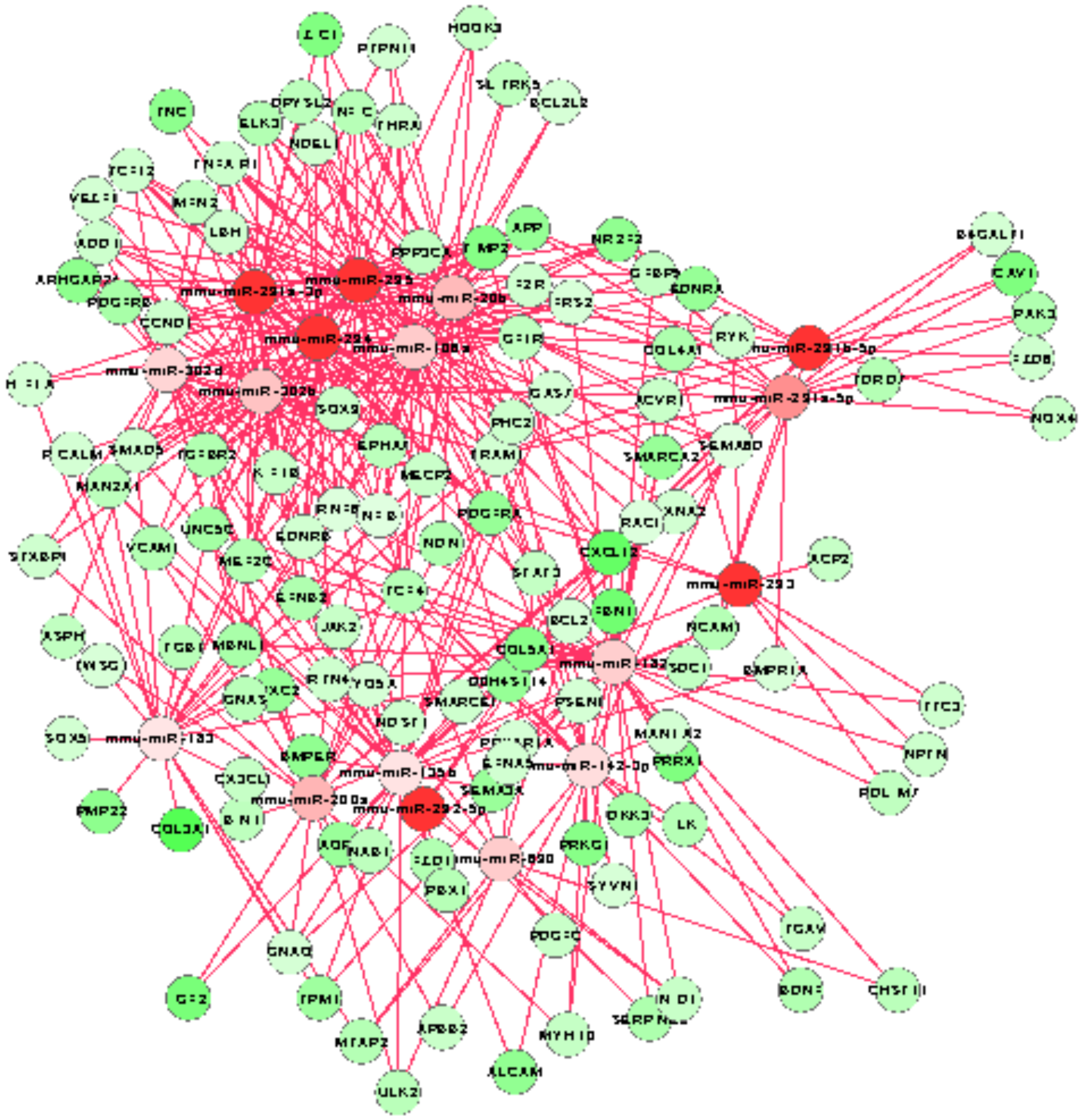

Down-regulated miRNAs mediate metabolism and pluripotency

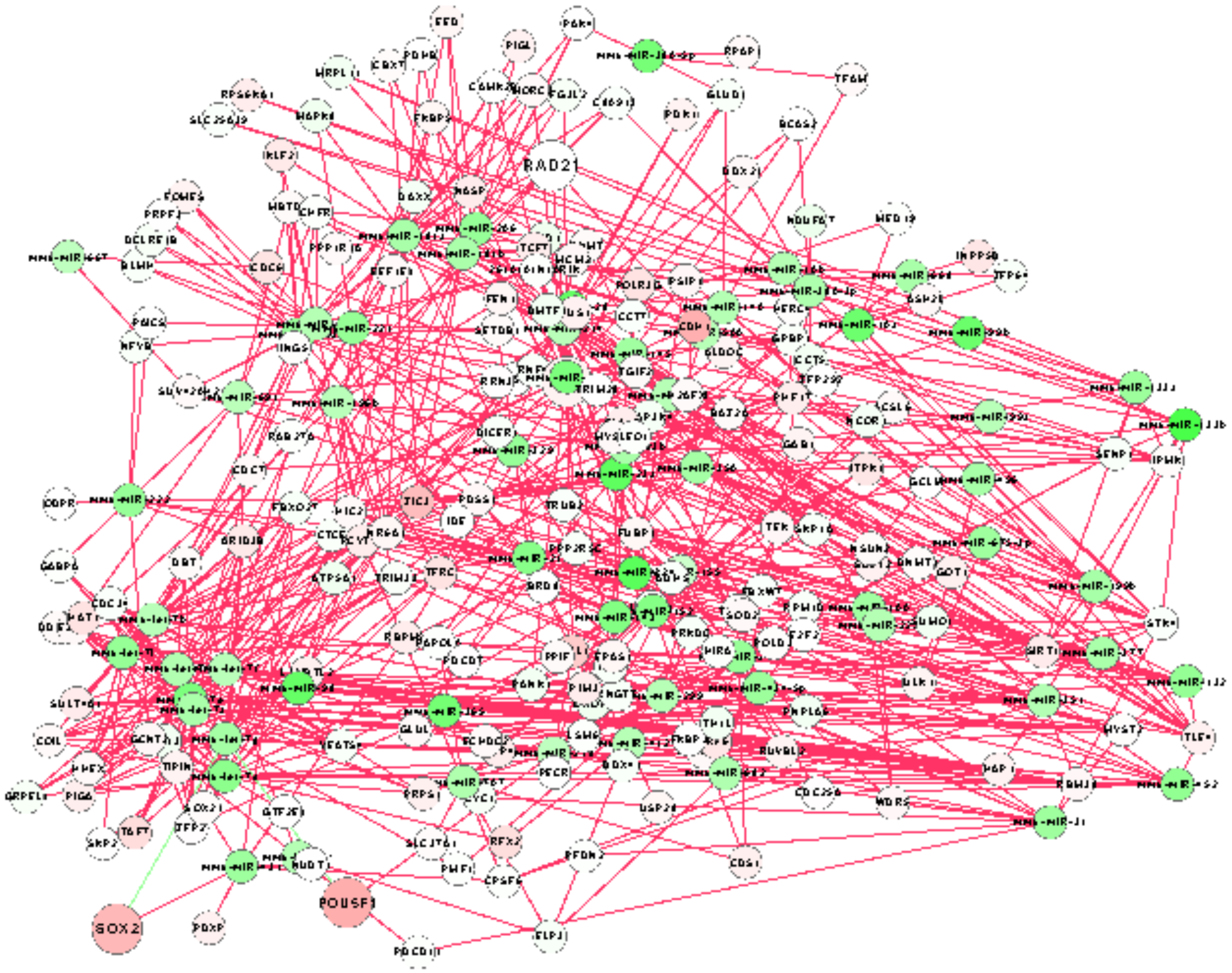

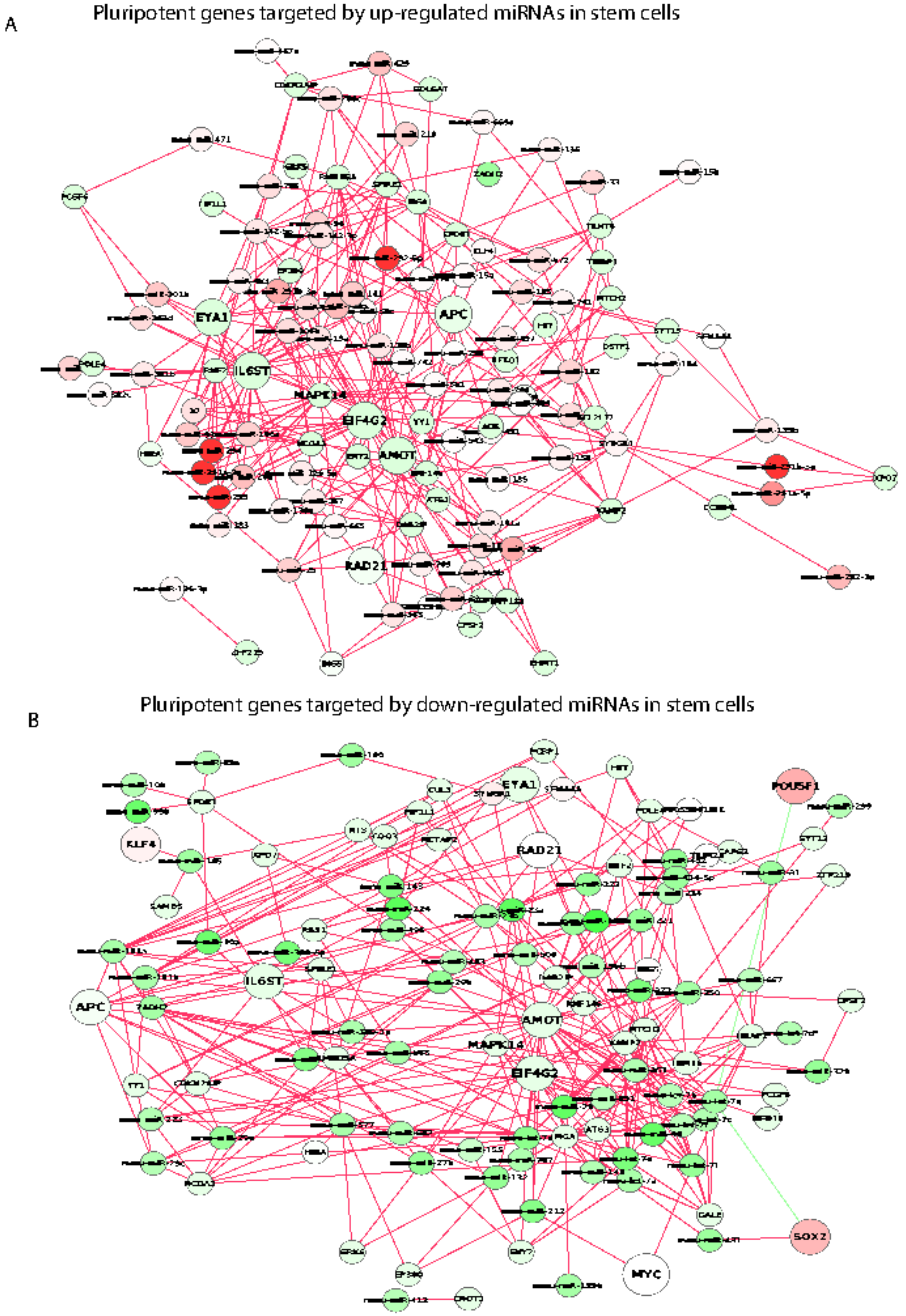
A
Pluripotent genes targeted by up-regulated miRNAs in stem cells
EYA1
APC
IL6ST
MAPK14
EIF4G2
AMOT
RAD21
B
Pluripotent genes targeted by down-regulated miRNAs in stem cells
POU5F1
KLF4
EYA1
RAD21
IL6ST
APC
AMOT
MAPK14
EIF4G2
SOX2
MYC

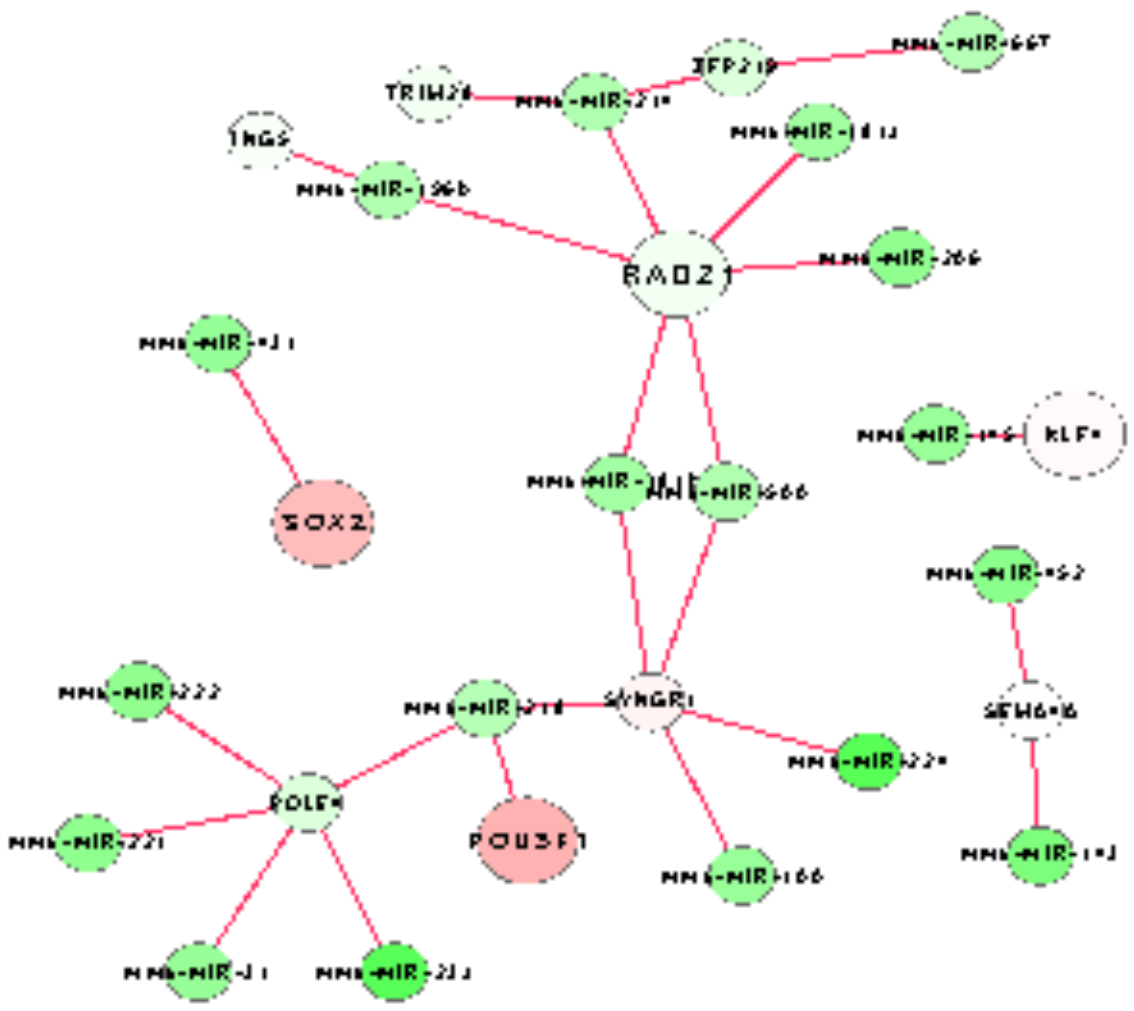
RAD21
SOX2
POU5F1

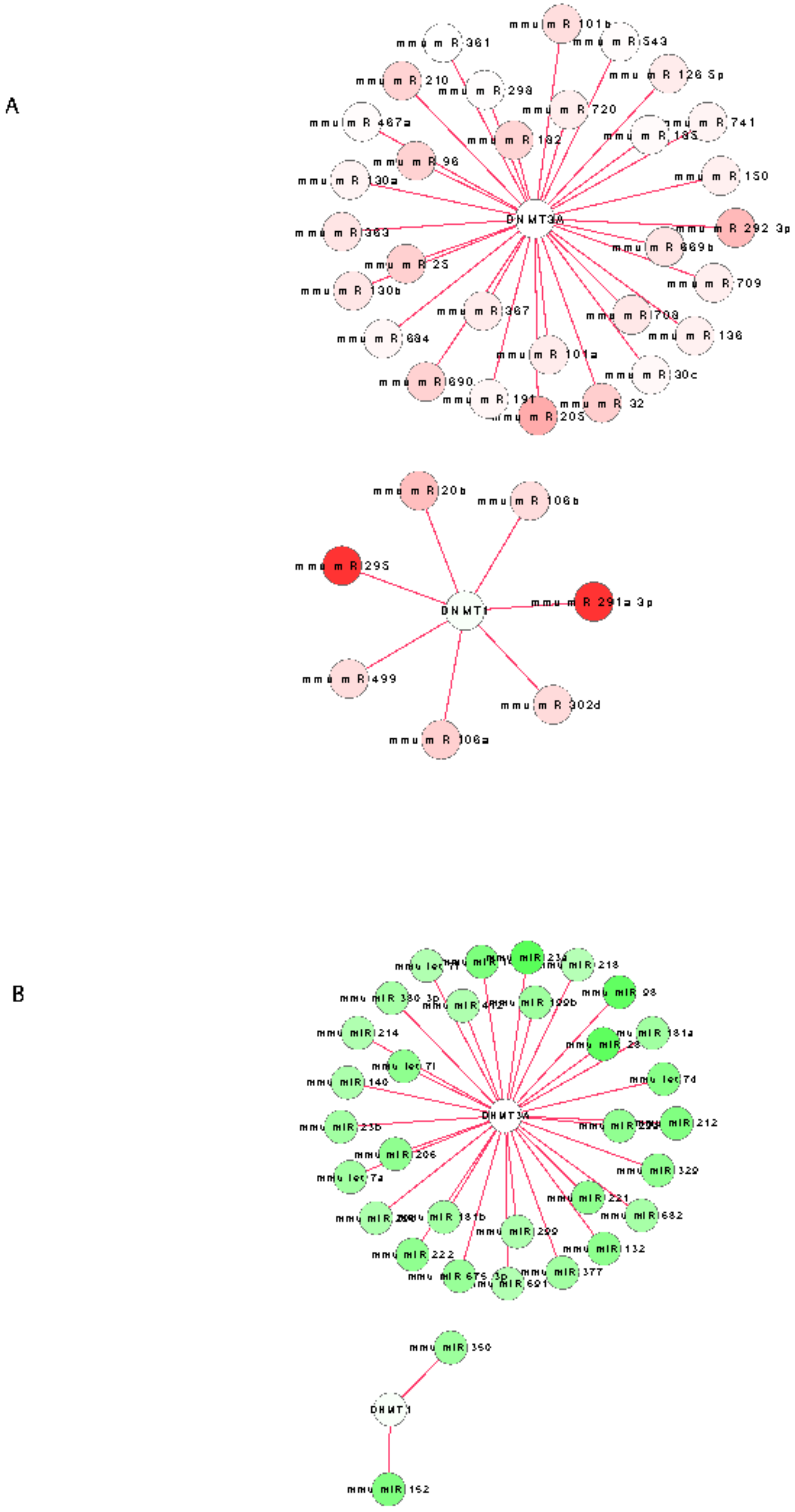

Table S1 data resources

| GEO number | Platform | measurement |
|---|---|---|
| GSE25310 | CLIP-seq | ES miRNA targets |
| GSE11724 | ChIP-seq | TF binding |
| GSE11431 | ChIP-seq | TF binding |
| GSM278905 | RRBS | ES DNA methylation |
| GSM278902 | RRBS | TKO methylation |
| GSM539867 | RNA-Seq | mouse embryonic fibroblast cells [09-002] |
| GSM539866 | RNA-Seq | mouse embryonic stem cells [09-002] |
| GSE30012 | Mouse430_2 array | ES miRNA expression |
| GSE25310 | GPL11410 | micrRNA and mRNA expression |
| GSM231739 | Agilent-GPL2872 | ES gene expression TKO/WT |
| GSE15267 | AffymetrixGPL1261 | Gene expression somatic cells vs stem cells |
| GSE14012 | AffymetrixGPL1261 | Gene expression somatic cells vs stem cells |
| GSE10871 | AffymetrixGPL1261 | Gene expression somatic cells vs stem cells |
| GSE14790 | AffymetrixGPL1261 | Gene expression somatic cells vs stem cells |
| GSE16062 | IlluminaGPL6885 | Gene expression somatic cells vs stem cells |

Table S2 miRNAs differentially expressed between stem cell vs somatic cells with high frequency >25% of observations

| geneID | foldChange.StemcellvsSomaticCells |
|---|---|
| mmu-miR-291a-3p | 23.38772103 |
| mmu-miR-292-5p | 20.9773225 |
| mmu-miR-295 | 18.61514609 |
| mmu-miR-290 | 16.83638073 |
| mmu-miR-291b-5p | 16.74397919 |
| mmu-miR-294 | 16.33979504 |
| mmu-miR-293 | 15.71944751 |
| mmu-miR-291a-5p | 10.70302558 |
| mmu-miR-291b-3p | 8.775662583 |
| mmu-miR-205 | 8.656636965 |
| mmu-miR-292-3p | 7.513040922 |
| mmu-miR-200a | 7.476828136 |
| mmu-miR-20b | 7.113430578 |
| mmu-miR-429 | 6.970376253 |
| mmu-miR-93 | 6.320621674 |
| mmu-miR-302b | 6.319533201 |
| mmu-miR-124a | 6.210366092 |
| mmu-miR-141 | 6.112532011 |
| mmu-miR-32 | 5.892597823 |
| mmu-miR-25 | 5.664093286 |
| mmu-miR-106a | 5.660199796 |
| mmu-miR-96 | 5.576097803 |
| mmu-miR-690 | 5.566673137 |
| mmu-miR-210 | 5.494180249 |
| mmu-miR-182 | 5.380304319 |
| mmu-miR-339 | 5.178218031 |
| mmu-miR-33 | 5.150748639 |
| mmu-miR-19b | 5.130661066 |
| mmu-miR-302d | 4.786597786 |
| mmu-miR-195 | 4.773097795 |
| mmu-miR-297b | 4.742119218 |
| mmu-miR-499 | 4.699375408 |
| mmu-miR-106b | 4.67924418 |
| mmu-miR-712* | 4.616388282 |
| mmu-miR-101b | 4.615452785 |
| mmu-miR-706 | 4.577798951 |
| mmu-miR-672 | 4.425689091 |
| mmu-miR-20a | 4.355849211 |
| mmu-miR-302 | 4.187271105 |
| mmu-miR-142-3p | 4.185529766 |
| mmu-miR-19a | 4.154314537 |
| mmu-miR-450b | 4.143615059 |
| mmu-miR-200c | 4.139625599 |
| mmu-miR-17-5p | 4.015734124 |
| mmu-miR-669b | 4.005643461 |
| mmu-miR-363 | 4.004376577 |
| mmu-miR-130b | 3.921190227 |
| mmu-miR-142-5p | 3.771559152 |
| mmu-miR-708 | 3.757978939 |
| mmu-miR-301b | 3.671137996 |
| mmu-miR-297 | 3.649760196 |
| mmu-miR-711 | 3.642487186 |
| mmu-miR-135b | 3.61287415 |
| mmu-miR-145 | -3.689012325 |
| mmu-miR-125b | -3.750522829 |
| mmu-miR-675-3p | -3.775797104 |
| mmu-miR-31 | -3.789955248 |
| mmu-miR-431 | -3.827836369 |
| mmu-miR-133a | -3.931634792 |
| mmu-miR-434-5p | -3.95970537 |
| mmu-miR-29a | -4.128390595 |
| mmu-miR-125a | -4.168574915 |
| mmu-miR-27b | -4.181204463 |
| mmu-miR-206 | -4.1991836 |
| mmu-miR-222 | -4.206742949 |
| mmu-miR-329 | -4.226853284 |
| mmu-let-7i | -4.253606445 |
| mmu-miR-221 | -4.277690509 |
| mmu-miR-132 | -4.327457901 |
| mmu-miR-345 | -4.362726416 |
| mmu-miR-452 | -4.493999108 |
| mmu-let-7d | -4.733310501 |
| mmu-miR-181a* | -4.890038681 |
| mmu-miR-212 | -5.03014549 |
| mmu-miR-152 | -5.091913825 |
| mmu-miR-143 | -5.251965264 |
| mmu-miR-433-5p | -5.353386842 |
| mmu-let-7e | -5.627257306 |
| mmu-miR-322 | -5.627467759 |
| mmu-miR-380-5p | -5.840198038 |
| mmu-miR-666 | -5.963478682 |
| mmu-miR-365 | -6.053925857 |
| mmu-miR-10a | -6.463178387 |
| mmu-miR-24* | -6.501407687 |
| mmu-miR-99b | -6.54057846 |
| mmu-miR-98 | -6.824098585 |
| mmu-miR-28 | -6.985496746 |
| mmu-miR-23a | -7.258185943 |
| mmu-miR-675-5p | -7.319121705 |
| mmu-miR-224 | -7.383165325 |
| mmu-miR-133b | -8.331863729 |
| mmu-miR-155 | -12.36089174 |

Table S3. Genes differentially expressed between stem cells and somatic cells with high frequency >50% of observations

| GB_ACC | ENTREZ_GI | Gene.Symbol | logFoldChange.ESvsSC |
|---|---|---|---|
| BC006640 | 20315 | Cxcl12 | -7.932773867 |
| BF225802 | 16011 | Igfbp5 | -7.7652649 |
| AW550625 | 12825 | Col3a1 | -7.7405454 |
| BF227507 | 12843 | Col1a2 | -7.50996995 |
| BB542051 | 18295 | Ogn | -7.13772715 |
| U08020 | 12842 | Col1a1 | -7.025342217 |
| NM_013655 | 20315 | Cxcl12 | -6.8464668 |
| NM_011581 | 21826 | Thbs2 | -6.743730283 |
| AV229424 | 12832 | Col5a2 | -6.6650371 |
| BB051738 | 18933 | Prrx1 | -6.523803517 |
| M65143 | 16948 | Lox | -6.263793183 |
| AF007248 | 14118 | Fbn1 | -6.234405867 |
| NM_009933 | 12833 | Col6a1 | -6.142877467 |
| NM_007993 | 14118 | Fbn1 | -6.087729233 |
| BC019502 | 12111 | Bgn | -6.042548083 |
| AB015978 | 18414 | Osmr | -6.018220117 |
| BE197945 | 20324 | Sdpr | -5.902171217 |
| NM_007729 | 12814 | Col11a1 | -5.888984717 |
| AW049660 | 18032 | Nfix | -5.878165733 |
| NM_010514 | 16002 | Igf2 | -5.805814817 |
| BB197591 | 54216 | Pcdh7 | -5.793524283 |
| AV226618 | 13837 | Epha3 | -5.703367667 |
| AF378762 | 69538 | Antxr1 | -5.671588117 |
| M68513 | 13837 | Epha3 | -5.646988233 |
| BC002064 | 19242 | Ptn | -5.631718783 |
| NM_011340 | 20317 | Serpinf1 | -5.561155483 |
| BB795075 | 18214 | Ddr2 | -5.5521091 |
| BC023060 | 216616 | Efemp1 | -5.541377517 |
| BF144658 | 21808 | Tgfb2 | -5.516599417 |
| BB250384 | 22329 | Vcam1 | -5.511377133 |
| AB029929 | 12389 | Cav1 | -5.478985417 |
| D67076 | 11504 | Adamts1 | -5.469247617 |
| BB315728 | 18032 | Nfix | -5.460392167 |
| BI794771 | 12842 | Col1a1 | -5.456734267 |
| L06502 | 18933 | Prrx1 | -5.448142767 |
| BG793483 | 21813 | Tgfbr2 | -5.434682533 |
| BB532202 | 14264 | Fmod | -5.394454433 |
| BB361162 | 22771 | Zic1 | -5.383117233 |
| AI931862 | 12111 | Bgn | -5.378983617 |
| BC014690 | 21809 | Tgfb3 | -5.302384367 |
| NM_018865 | 22402 | Wisp1 | -5.29682725 |
| AV064339 | 20324 | Sdpr | -5.2953667 |
| BG067986 | 54216 | Pcdh7 | -5.295023283 |
| NM_030888 | 81799 | C1qtnf3 | -5.279631533 |
| BB233297 | 21826 | Thbs2 | -5.26170495 |
| BF168458 | 21858 | Timp2 | -5.239630733 |
| BB259670 | 14362 | Fzd1 | -5.23704885 |
| AA499047 | 18596 | Pdgfrb | -5.226616983 |
| BC026446 | 219151 | Scara3 | -5.1983228 |
| NM_010714 | 16876 | Lhx9 | -5.18046395 |
| NM_133859 | 99543 | Olfml3 | -5.167943083 |
| BB067079 | 22418 | Wnt5a | -5.160633383 |
| AW555664 | 399558 | Flrt2 | -5.14692565 |
| BC021484 | 16651 | Sspn | -5.141520367 |
| BC011063 | 15402 | Hoxa5 | -5.109899467 |
| M93954 | 21858 | Timp2 | -5.0922978 |
| AV310588 | 12444 | Ccnd2 | -5.076240083 |
| NM_013589 | 16997 | Ltbp2 | -5.07117345 |
| BB791906 | 14178 | Fgf7 | -5.0607749 |
| BQ175880 | 12444 | Ccnd2 | -5.031714083 |
| NM_009829 | 12444 | Ccnd2 | -5.03108855 |
| BC022107 | 12558 | Cdh2 | -5.030543317 |
| NM_011160 | 19091 | Prkg1 | -5.010924967 |
| AK007904 | 12444 | Ccnd2 | -5.010597567 |
| BB114398 | 12816 | Col12a1 | -5.001727783 |
| BB377340 | 13612 | Edil3 | -4.984431667 |
| BI455189 | 12834 | Col6a2 | -4.981937467 |
| BC018425 | 22418 | Wnt5a | -4.949481167 |
| AF309564 | 80837 | Rhoj | -4.942552217 |
| AW744319 | 12831 | Col5a1 | -4.912884333 |
| NM_010284 | 14600 | Ghr | -4.907539267 |
| AK003674 | 68588 | Cthrc1 | -4.9040982 |
| NM_009610 | 11468 | Actg2 | -4.88703015 |
| BM220945 | 239217 | Kctd12 | -4.885967567 |
| BB658835 | 23794 | Adamts5 | -4.884421383 |
| BC019836 | 16010 | Igfbp4 | -4.86110935 |
| NM_018764 | 54216 | Pcdh7 | -4.826503217 |
| AA500897 | 26362 | Axl | -4.824049367 |
| AV222756 | 259300 | Ehd2 | -4.811690783 |
| NM_018884 | 55983 | Pdzrn3 | -4.770790783 |
| BB759833 | 17300 | Foxc1 | -4.766068317 |
| BQ177170 | 17112 | Tm4sf1 | -4.749287983 |
| BM250666 | 12830 | Col4a5 | -4.738634083 |
| BC025502 | 231532 | Arhgap24 | -4.7154329 |
| NM_007392 | 11475 | Acta2 | -4.71167825 |
| NM_008885 | 18858 | Pmp22 | -4.71162355 |
| AV246911 | 12831 | Col5a1 | -4.70608125 |
| NM_133918 | 100952 | Emilin1 | -4.685038 |
| AV260198 | 83675 | Bicc1 | -4.676352617 |
| BC021376 | 225207 | Zfp521 | -4.6747989 |
| NM_011607 | 21923 | Tnc | -4.6736706 |
| NM_007833 | 13179 | Dcn | -4.668888083 |
| AF357006 | 16949 | Loxl1 | -4.661639267 |
| BC011507 | 74480 | Samd4 | -4.6606821 |
| BB493031 | 23888 | Gpc6 | -4.659793267 |
| BI249259 | 18032 | Nfix | -4.659746767 |
| NM_054044 | 78560 | Gpr124 | -4.652361 |
| BC001999 | 13732 | Emp3 | -4.648468717 |
| BB041180 | 16876 | Lhx9 | -4.631340583 |
| BI690209 | 57342 | Parva | -4.603038383 |
| AK018128 | 70717 | 6330406I15Ri | -4.599817333 |
| AI595932 | 17260 | Mef2c | -4.571089667 |
| AI463873 | 11819 | Nr2f2 | -4.558937633 |
| BG075699 | 399558 | Flrt2 | -4.550595967 |
| AV149705 | 108927 | Lhfp | -4.537199083 |
| AK014221 | 73230 | Bmper | -4.536276183 |
| BB468025 | 54381 | Pgcp | -4.535907333 |
| BB363812 | 14051 | Eya4 | -4.52926855 |
| AK020118 | 23888 | Gpc6 | -4.5247486 |
| BM217996 | 83675 | Bicc1 | -4.521218667 |
| NM_009636 | 11568 | Aebp1 | -4.509546617 |
| AA987181 | 15405 | Hoxa9 | -4.496607467 |
| NM_011766 | 22762 | Zfpm2 | -4.496262983 |
| AV315205 | 11658 | Alcam | -4.495887367 |
| U95030 | 11658 | Alcam | -4.4742135 |
| BF147716 | 17390 | Mmp2 | -4.463715417 |
| NM_008608 | 17387 | Mmp14 | -4.4584154 |
| AV332957 | 20429 | Shox2 | -4.44107 |
| AK008112 | 18933 | Prrx1 | -4.429082483 |
| BB787243 | 16010 | Igfbp4 | -4.408829283 |
| M27130 | 12505 | Cd44 | -4.408205433 |
| NM_023118 | 13132 | Dab2 | -4.401669267 |
| BB817332 | 399558 | Flrt2 | -4.401003567 |
| D63423 | 11747 | Anxa5 | -4.3868292 |
| BC014870 | 235587 | Parp3 | -4.368841483 |
| U03425 | 13649 | Egfr | -4.353990467 |
| NM_011150 | 19039 | Lgals3bp | -4.34716645 |
| BB229377 | 13717 | Eln | -4.3469066 |
| BF681826 | 26360 | Angptl2 | -4.34243245 |
| AV369812 | 13649 | Egfr | -4.340586717 |
| AI385532 | 21825 | Thbs1 | -4.3388455 |
| AK012411 | 20317 | Serpinf1 | -4.334218667 |
| NM_016900 | 12390 | Cav2 | -4.327267733 |
| NM_007471 | 11820 | App | -4.32309575 |
| BC026153 | 13841 | Epha7 | -4.322886683 |
| BE652876 | 60527 | Fads3 | -4.31762715 |
| NM_022563 | 18214 | Ddr2 | -4.317137833 |
| BC005490 | 11820 | App | -4.311839833 |
| AY083458 | 235505 | Cd109 | -4.310650267 |
| AW146109 | 12505 | Cd44 | -4.309742383 |
| AF022889 | 268977 | Ltbp1 | -4.293317933 |
| BF780807 | 23972 | Papss2 | -4.2837158 |
| NM_010222 | 14231 | Fkbp7 | -4.283483917 |
| BB765827 | 74761 | Mxra8 | -4.28266945 |
| BB369191 | 27528 | D0H4S114 | -4.258134683 |
| NM_011658 | 22160 | Twist1 | -4.252172233 |
| BC022679 | 52552 | Parp8 | -4.249244683 |
| BC024375 | 14600 | Ghr | -4.235411317 |
| AW558570 | 13617 | Ednra | -4.235029783 |
| NM_010151 | 13865 | Nr2f1 | -4.23266835 |
| BB053506 | 207181 | Rbms3 | -4.227119367 |
| BB464523 | 12293 | Cacna2d1 | -4.2176231 |
| AI325255 | 14230 | Fkbp10 | -4.2139772 |
| NM_009365 | 21804 | Tgfb1i1 | -4.2037574 |
| AW537708 | 18595 | Pdgfra | -4.201673433 |
| BG244279 | 26360 | Angptl2 | -4.19122825 |
| AV021105 | 208647 | Creb3l2 | -4.189322833 |
| NM_021474 | 58859 | Efemp2 | -4.172999333 |
| AW558468 | 230103 | Npr2 | -4.170204383 |
| BM230959 | 229731 | Slc25a24 | -4.15558595 |
| NM_013519 | 14234 | Foxc2 | -4.150217817 |
| BG248060 | 12153 | Bmp1 | -4.1456309 |
| AK011935 | 67155 | Smarca2 | -4.1424534 |
| BC009660 | 109042 | Prkcdbp | -4.138494033 |
| NM_010730 | 16952 | Anxa1 | -4.137948167 |
| NM_017464 | 18003 | Nedd9 | -4.125438817 |
| NM_020606 | 57342 | Parva | -4.117250083 |
| NM_054042 | 70445 | Cd248 | -4.1162995 |
| AW107196 | 74480 | Samd4 | -4.089650683 |
| NM_008520 | 16998 | Ltbp3 | -4.088398567 |
| NM_007400 | 11489 | Adam12 | -4.087465183 |
| BB138485 | 70676 | Gulp1 | -4.06900915 |
| BB041237 | 17389 | Mmp16 | -4.0669223 |
| NM_009152 | 20346 | Sema3a | -4.06663545 |
| AK007400 | 77889 | Lbh | -4.065124233 |
| BC027199 | 233328 | Lrrk1 | -4.0638856 |
| BC003726 | 14789 | Leprel2 | -4.061919033 |
| BM220576 | 53623 | Gria3 | -4.059143167 |
| BM570006 | 20564 | Slit3 | -4.050010933 |
| NM_007472 | 11826 | Aqp1 | -4.049342233 |
| BB329489 | 12406 | Serpinh1 | -4.035949167 |
| BC024358 | 22004 | Tpm2 | -4.027799067 |
| BB371406 | 57265 | Fzd2 | -4.024201683 |
| BI106777 | 12293 | Cacna2d1 | -4.016322467 |
| BB484759 | 224024 | Scarf2 | -3.995337517 |
| BC025145 | 19266 | Ptprd | -3.96890555 |
| AK004519 | 56726 | Sh3bgrl | -3.915095233 |
| NM_010221 | 14230 | Fkbp10 | -3.911402667 |
| BB114067 | 21345 | Tagln | -3.906768233 |
| NM_010517 | 16010 | Igfbp4 | -3.895716167 |
| AV164956 | 22418 | Wnt5a | -3.889125017 |
| U71189 | 13592 | Ebf2 | -3.883624733 |
| AF339910 | 18605 | Enpp1 | -3.88293505 |
| BQ173967 | 13640 | Efna5 | -3.882553467 |
| AV246882 | 235505 | Cd109 | -3.87793775 |
| AK003186 | 22004 | Tpm2 | -3.875057833 |
| NM_009821 | 12394 | Runx1 | -3.8666507 |
| NM_010743 | 17082 | Il1rl1 | -3.865813883 |
| BB324823 | 16998 | Ltbp3 | -3.86059405 |
| BC025514 | 102644 | Oaf | -3.857056783 |
| NM_007801 | 13036 | Ctsh | -3.85513275 |
| NM_010581 | 16423 | Cd47 | -3.854950117 |
| AV238225 | 16905 | Lmna | -3.8510058 |
| AK018679 | 16423 | Cd47 | -3.832084267 |
| BG073383 | 15410 | Hoxb3 | -3.820933467 |
| NM_026405 | 67844 | Rab32 | -3.819808667 |
| BF144687 | 70676 | Gulp1 | -3.81964405 |
| NM_019391 | 16985 | Lsp1 | -3.81878645 |
| NM_013586 | 16950 | Loxl3 | -3.813466317 |
| AI415741 | 94352 | Loxl2 | -3.8120326 |
| BI220012 | 12406 | Serpinh1 | -3.811311317 |
| NM_007802 | 13038 | Ctsk | -3.804638867 |
| BB832504 | 18028 | Nfib | -3.801083117 |
| BC013463 | 15430 | Hoxd10 | -3.784799 |
| M22479 | 22003 | Tpm1 | -3.783677167 |
| BG963150 | 20563 | Slit2 | -3.783541217 |
| BC013560 | 12827 | Col4a2 | -3.774358167 |
| NM_008984 | 19274 | Ptprm | -3.769132717 |
| BB233088 | 21928 | Tnfaip2 | -3.7684148 |
| BC008277 | 13617 | Ednra | -3.767968717 |
| BI111620 | 21859 | Timp3 | -3.76779495 |
| BB475194 | 23794 | Adamts5 | -3.76137165 |
| NM_021355 | 14264 | Fmod | -3.736082733 |
| NM_007585 | 12306 | Anxa2 | -3.731015333 |
| BF578055 | 229731 | Slc25a24 | -3.725295933 |
| BB547877 | 20563 | Slit2 | -3.716729817 |
| NM_016753 | 17035 | Lxn | -3.70538055 |
| AV016275 | 20742 | Spnb2 | -3.701173833 |
| NM_008495 | 16852 | Lgals1 | -3.6991015 |
| AW551930 | 83675 | Bicc1 | -3.698414167 |
| AK018466 | 207181 | Rbms3 | -3.68959625 |
| NM_008987 | 19288 | Ptx3 | -3.687359367 |
| AK003819 | 74777 | Sepn1 | -3.6822946 |
| AW412729 | 12816 | Col12a1 | -3.663389833 |
| BC020152 | 223272 | Itgbl1 | -3.657189217 |
| Y07687 | 18028 | Nfib | -3.657044833 |
| NM_011985 | 26561 | Mmp23 | -3.644890267 |
| NM_009378 | 21824 | Thbd | -3.6371471 |
| BF451748 | 107589 | Mylk | -3.626035767 |
| BM251152 | 13003 | Vcan | -3.620597833 |
| BB468082 | 18481 | Pak3 | -3.6195631 |
| NM_007899 | 13601 | Ecm1 | -3.6077534 |
| BB522674 | 20541 | Slc8a1 | -3.606644633 |
| NM_019989 | 56726 | Sh3bgrl | -3.60465905 |
| BC023448 | 23888 | Gpc6 | -3.603901967 |
| AK002516 | 11857 | Arhgdib | -3.6024453 |
| BF100813 | 13618 | Ednrb | -3.594391017 |
| NM_011777 | 22793 | Zyx | -3.594158817 |
| BB326929 | 58194 | Sh3kbp1 | -3.582396833 |
| BB248904 | 56726 | Sh3bgrl | -3.58152235 |
| BB027759 | 231997 | Fkbp14 | -3.5728165 |
| AU021035 | 15529 | Sdc2 | -3.570693117 |
| AI481026 | 29817 | Igfbp7 | -3.56210855 |
| NM_008393 | 16373 | Irx3 | -3.560717517 |
| BB526042 | 320452 | P4ha3 | -3.5487859 |
| BB100920 | 17118 | Marcks | -3.53907495 |
| BB097480 | 223254 | Farp1 | -3.534434933 |
| BI687652 | 18028 | Nfib | -3.524464033 |
| BG868949 | 17122 | Mxd4 | -3.514140767 |
| NM_020561 | 57319 | Smpdl3a | -3.51306225 |
| AF059567 | 12579 | Cdkn2b | -3.5073857 |
| NM_009472 | 22253 | Unc5c | -3.505214767 |
| NM_009794 | 12334 | Capn2 | -3.498667433 |
| NM_009627 | 11535 | Adm | -3.48734415 |
| BQ175796 | 18481 | Pak3 | -3.4733871 |
| BC005569 | 58809 | Rnase4 | -3.460646467 |
| BB327018 | 14118 | Fbn1 | -3.456580617 |
| NM_010212 | 14200 | Fhl2 | -3.453920533 |
| BB313689 | 20541 | Slc8a1 | -3.452686383 |
| AK019164 | 17475 | Mpdz | -3.437966567 |
| AW702161 | 11745 | Anxa3 | -3.4355384 |
| BC027084 | 259300 | Ehd2 | -3.43459445 |
| BC005686 | 13713 | Elk3 | -3.43373535 |
| BB065799 | 240725 | Sulf1 | -3.429931133 |
| NM_130449 | 140792 | Colec12 | -3.424621383 |
| BB823350 | 19092 | Prkg2 | -3.412943967 |
| AI642438 | 16852 | Lgals1 | -3.4127125 |
| NM_007855 | 13345 | Twist2 | -3.412527233 |
| BB787292 | 14302 | Frk | -3.411315533 |
| BB354684 | 217410 | Trib2 | -3.410151567 |
| AF117951 | 94352 | Loxl2 | -3.400751667 |
| BC025897 | 225791 | Zadh2 | -3.398803383 |
| NM_019877 | 56358 | Copz2 | -3.386908433 |
| AV162270 | 22240 | Dpysl3 | -3.3799255 |
| D63383 | 20465 | Sim2 | -3.378190883 |
| NM_008546 | 17150 | Mfap2 | -3.375675317 |
| NM_008008 | 14178 | Fgf7 | -3.373739283 |
| BQ174721 | 214791 | Sertad4 | -3.37195165 |
| BB377873 | 121021 | Cspg4 | -3.368490867 |
| BB542535 | 13036 | Ctsh | -3.3666448 |
| BC024876 | 223453 | Dap | -3.364419667 |
| L24755 | 12153 | Bmp1 | -3.360524783 |
| X66083 | 12505 | Cd44 | -3.354735533 |
| AW547821 | 17268 | Meis1 | -3.3441435 |
| BC016447 | 26427 | Creb3l1 | -3.342256433 |
| BM213516 | 20742 | Spnb2 | -3.33549655 |
| BM220880 | 108655 | Foxp1 | -3.326646233 |
| AK004853 | 50781 | Dkk3 | -3.314331467 |
| NM_008809 | 18596 | Pdgfrb | -3.31185265 |
| AV245241 | 259300 | Ehd2 | -3.304100233 |
| NM_020007 | 56758 | Mbnl1 | -3.289487817 |
| BC028307 | 68792 | Srpx2 | -3.28413355 |
| AA709993 | 20365 | Serf1 | -3.275465433 |
| BC006737 | 12370 | Casp8 | -3.272708067 |
| BB151715 | 23871 | Ets1 | -3.271100167 |
| BI452727 | 14314 | Fstl1 | -3.26508275 |
| BB003393 | 21923 | Tnc | -3.2549266 |
| BB535494 | 18003 | Nedd9 | -3.2516385 |
| BB779859 | 209448 | Hoxc10 | -3.2440814 |
| BI248947 | 109624 | Cald1 | -3.241609283 |
| L26349 | 21937 | Tnfrsf1a | -3.2395265 |
| BC012674 | 19285 | Ptrf | -3.228721367 |
| U30244 | 13642 | Efnb2 | -3.225253617 |
| AK004598 | 71703 | Armcx3 | -3.218577033 |
| BB075247 | 19266 | Ptprd | -3.217064717 |
| BB766878 | 107581 | Col16a1 | -3.216479983 |
| NM_019586 | 56228 | Ube2j1 | -3.216275217 |
| NM_008761 | 18301 | Fxyd5 | -3.21540255 |
| M24849 | 22352 | Vim | -3.212612833 |
| BB366804 | 18230 | Nxn | -3.208821517 |
| AV031691 | 22771 | Zic1 | -3.207325183 |
| AK003303 | 19266 | Ptprd | -3.1900006 |
| NM_009252 | 20716 | Serpina3n | -3.1897822 |
| BM233698 | 12977 | Csf1 | -3.186553767 |
| AK018666 | 50766 | Crim1 | -3.184091083 |
| NM_013813 | 13823 | Epb4.1l3 | -3.17559455 |
| BB097063 | 16826 | Ldb2 | -3.167842717 |
| BG070361 | 18514 | Pbx1 | -3.16060585 |
| NM_023564 | 70310 | Plscr3 | -3.156913317 |
| BC016893 | 15414 | Hoxb6 | -3.149439067 |
| BC025602 | 52377 | Rcn3 | -3.143586083 |
| BE628275 | 68910 | Zfp467 | -3.129177833 |
| BB751459 | 240725 | Sulf1 | -3.118822083 |
| NM_023813 | 108058 | Camk2d | -3.106603767 |
| BE630020 | 19122 | Prnp | -3.099244017 |
| NM_015786 | 50708 | Hist1h1c | -3.096520833 |
| NM_010736 | 17000 | Ltbr | -3.096244717 |
| NM_008380 | 16323 | Inhba | -3.090835667 |
| AF080090 | 20350 | Sema3f | -3.086323 |
| AV220340 | 229534 | Pbxip1 | -3.08320105 |
| NM_016846 | 19731 | Rgl1 | -3.083172067 |
| AA510713 | 56468 | Socs5 | -3.08271985 |
| AV246759 | 21808 | Tgfb2 | -3.078846817 |
| AV238718 | 233733 | Galntl4 | -3.07739925 |
| BB311061 | 56876 | Nelf | -3.07710745 |
| BM231794 | 216188 | Aldh1l2 | -3.071789317 |
| NM_053147 | 93893 | Pcdhb22 | -3.06913275 |
| AV124445 | 17984 | Ndn | -3.067003683 |
| AI747133 | 12338 | Capn6 | -3.061495833 |
| NM_009784 | 12293 | Cacna2d1 | -3.0553154 |
| NM_008675 | 17965 | Nbl1 | -3.054173417 |
| X58380 | 15364 | Hmga2 | -3.053552917 |
| NM_009468 | 22240 | Dpysl3 | -3.0489959 |
| NM_019390 | 16905 | Lmna | -3.043480833 |
| BB533903 | 50708 | Hist1h1c | -3.028666667 |
| BB250811 | 18542 | Pcolce | -3.026433 |
| BE951265 | 66180 | 1110036O03R | -3.02614895 |
| AW546141 | 17118 | Marcks | -3.021924833 |
| BB825801 | 13713 | Elk3 | -3.021571583 |
| NM_009255 | 20720 | Serpine2 | -3.02130325 |
| BF158638 | 12826 | Col4a1 | -3.010500317 |
| NM_016886 | 53623 | Gria3 | -3.003217433 |
| NM_009925 | 12813 | Col10a1 | -3.001262167 |
| AV278675 | 244448 | Triml1 | 3.01398255 |
| BG966751 | 104156 | Etv5 | 3.0156244 |
| X67668 | 97165 | Hmgb2 | 3.020431667 |
| BB089717 | 12449 | Ccnf | 3.022975367 |
| BB292220 | 58180 | Hic2 | 3.025155533 |
| BG069191 | 50927 | Nasp | 3.027539567 |
| BG066754 | 67204 | Eif2s2 | 3.0295477 |
| AK019115 | 68239 | Krt42 | 3.03983065 |
| AU015121 | 268697 | Ccnb1 | 3.042294933 |
| NM_008566 | 17218 | Mcm5 | 3.058338333 |
| NM_019812 | 93759 | Sirt1 | 3.064211417 |
| NM_020599 | 19771 | Rlbp1 | 3.06822615 |
| AV280841 | 219132 | D14Ertd668e | 3.078624667 |
| BB493242 | 50927 | Nasp | 3.081964833 |
| NM_022409 | 63872 | Zfp296 | 3.082454667 |
| NM_013873 | 29859 | Sult4a1 | 3.0839579 |
| AF302127 | 72388 | Ripk4 | 3.086288367 |
| NM_011597 | 21873 | Tjp2 | 3.087218833 |
| BF578266 | 227327 | B3gnt7 | 3.0910465 |
| AI195532 | 216456 | Gls2 | 3.091450267 |
| BE650359 | 13619 | Phc1 | 3.103443667 |
| BM247863 | 56380 | Arid3b | 3.10414725 |
| BM935811 | 16403 | Itga6 | 3.109543333 |
| BG071670 | 75796 | Cdyl2 | 3.110318667 |
| BC006704 | 226419 | Dyrk3 | 3.114882817 |
| NM_008628 | 17685 | Msh2 | 3.118807583 |
| BM196098 | 14562 | Gdf3 | 3.129241967 |
| NM_020567 | 57441 | Gmnn | 3.129819017 |
| AK011289 | 66953 | Cdca7 | 3.133227817 |
| NM_009426 | 22044 | Trh | 3.133924383 |
| BB367422 | 239759 | Liph | 3.1340973 |
| BC011474 | 16818 | Lck | 3.140572717 |
| NM_011176 | 19143 | St14 | 3.14224885 |
| BB667216 | 22371 | Vwf | 3.143579333 |
| BI739053 | 12725 | Clcn3 | 3.159571817 |
| NM_011487 | 20849 | Stat4 | 3.169867333 |
| BQ177140 | 218038 | Amph | 3.17188915 |
| NM_008452 | 16598 | Klf2 | 3.1728325 |
| AF329833 | 66790 | Grtp1 | 3.203848083 |
| BB540053 | 217653 | C79407 | 3.221807417 |
| NM_007691 | 12649 | Chek1 | 3.229280333 |
| BC016095 | 228421 | Kif18a | 3.231334583 |
| BC017621 | 223775 | Pim3 | 3.23674225 |
| U42190 | 17688 | Msh6 | 3.242437417 |
| NM_027288 | 110173 | Manba | 3.25011815 |
| BE951628 | 217653 | C79407 | 3.269844583 |
| BB125424 | 242894 | Actr3b | 3.270038817 |
| AK010648 | 70024 | Mcm10 | 3.288071933 |
| AK006509 | 75796 | Cdyl2 | 3.304674333 |
| NM_010174 | 14077 | Fabp3 | 3.309141333 |
| NM_053271 | 116838 | Rims2 | 3.314384433 |
| BG070553 | 110957 | D1Pas1 | 3.318172517 |
| NM_133664 | 16763 | Lad1 | 3.323882083 |
| BQ176661 | 54484 | Mkrn1 | 3.3289175 |
| AF237702 | 20425 | Shmt1 | 3.331763533 |
| NM_016851 | 54139 | Irf6 | 3.333859083 |
| AA543734 | 21386 | Tbx3 | 3.333990817 |
| NM_010346 | 14786 | Grb7 | 3.34545575 |
| BC016571 | 50799 | Slc25a13 | 3.36053825 |
| AK007201 | 14536 | Nr6a1 | 3.360795183 |
| NM_013514 | 13829 | Epb4.9 | 3.361635267 |
| U39203 | 16331 | Inpp5d | 3.380223367 |
| AW552407 | 217837 | Itpk1 | 3.380479 |
| AK011462 | 72148 | 2610019F03R | 3.381959433 |
| BB397151 | 332221 | Zscan10 | 3.400120667 |
| U31625 | 12189 | Brca1 | 3.4060535 |
| BM225706 | 243574 | Kbtbd8 | 3.415306283 |
| NM_031386 | 83560 | Tex14 | 3.416736917 |
| NM_009171 | 20425 | Shmt1 | 3.449195983 |
| AK011986 | 112422 | 2610305D13R | 3.475925817 |
| AV305746 | 231327 | Ppat | 3.479371517 |
| BI465857 | 12224 | Klf5 | 3.483464667 |
| NM_021308 | 57746 | Piwil2 | 3.4897425 |
| BB758095 | 21873 | Tjp2 | 3.499096167 |
| NM_016964 | 50878 | Stag3 | 3.503349717 |
| BG066919 | 17765 | Mtf2 | 3.526635767 |
| BE986504 | 71801 | Plekhf2 | 3.556072867 |
| AV277242 | 243963 | Zfp473 | 3.576478483 |
| NM_011799 | 23834 | Cdc6 | 3.57881495 |
| NM_026190 | 67486 | Polr3g | 3.6006572 |
| NM_010357 | 14860 | Gsta4 | 3.6286395 |
| AU016127 | 67374 | Jam2 | 3.629554333 |
| BM230524 | 217944 | Rapgef5 | 3.6361713 |
| NM_022024 | 63986 | Gmfg | 3.640558983 |
| AK014330 | 107435 | Hat1 | 3.648040833 |
| NM_008816 | 18613 | Pecam1 | 3.660811183 |
| BC003329 | 54484 | Mkrn1 | 3.66147285 |
| AK011596 | 22042 | Tfrc | 3.664959917 |
| BB128741 | 319801 | 9630033F20R | 3.701341 |
| AF037370 | 12865 | Cox7a1 | 3.709307883 |
| BC006646 | 12224 | Klf5 | 3.738650033 |
| AV024881 | 52609 | Cbx7 | 3.776539367 |
| BB234337 | 101602 | AI467606 | 3.805827767 |
| AK013117 | 76974 | 1190003J15Ri | 3.806894467 |
| NM_010068 | 13436 | Dnmt3b | 3.83760305 |
| NM_009886 | 12614 | Celsr1 | 3.862550233 |
| BE648070 | 58180 | Hic2 | 3.868284683 |
| NM_007469 | 11812 | Apoc1 | 3.873187833 |
| BC018383 | 226180 | Ina | 3.88687835 |
| AB072269 | 13511 | Dsg2 | 3.8942605 |
| NM_013512 | 13824 | Epb4.1l4a | 3.902842033 |
| BB473446 | 239157 | Pnma2 | 3.946305617 |
| AK010391 | 71988 | Esco2 | 3.965438167 |
| BC020144 | 13511 | Dsg2 | 3.9713275 |
| AK009960 | 69697 | 2310057J16Ri | 3.9715563 |
| AF285580 | 77485 | Stk31 | 3.971668 |
| NM_007397 | 11481 | Acvr2b | 3.974705033 |
| BC003778 | 21420 | Tcfap2c | 4.010026117 |
| NM_011934 | 26380 | Esrrb | 4.057861333 |
| NM_016907 | 20732 | Spint1 | 4.058743833 |
| BC004654 | 19725 | Rfx2 | 4.076663867 |
| AV060866 | 18778 | Pla2g1b | 4.077706683 |
| BE133749 | 54484 | Mkrn1 | 4.082165 |
| AF425084 | 97848 | Serpinb6c | 4.1037625 |
| AV213552 | 24074 | Taf7 | 4.123073 |
| BQ175337 | 72685 | Dnajc6 | 4.13664 |
| AF104416 | 11828 | Aqp3 | 4.179715783 |
| BM210256 | 14106 | Foxh1 | 4.211211117 |
| NM_023844 | 67374 | Jam2 | 4.2200985 |
| AK019319 | 11816 | Apoe | 4.313206417 |
| NM_031383 | 83557 | Lin28 | 4.338394817 |
| BG084230 | 66824 | Pycard | 4.33866605 |
| AK010826 | 67374 | Jam2 | 4.358034267 |
| AF330212 | 12310 | Calca | 4.358796767 |
| BQ173923 | 12808 | Cobl | 4.384203133 |
| M18775 | 17762 | Mapt | 4.42236835 |
| BC027285 | 68713 | Ifitm1 | 4.4224295 |
| AK021186 | 77485 | Stk31 | 4.436390117 |
| NM_009903 | 12740 | Cldn4 | 4.437383 |
| NM_009749 | 12069 | Bex2 | 4.45785005 |
| NM_009434 | 22113 | Phlda2 | 4.474681333 |
| BC017609 | 18424 | Otx2 | 4.492426433 |
| BC023498 | 72027 | Slc39a4 | 4.528567183 |
| AA215046 | 67080 | 1700019D03R | 4.535403217 |
| BC004037 | 22256 | Ung | 4.580378817 |
| NM_007515 | 11989 | Slc7a3 | 4.595658 |
| BB469300 | 243574 | Kbtbd8 | 4.605572983 |
| NM_016754 | 17907 | Mylpf | 4.613079 |
| AV277440 | 18020 | Nfatc2ip | 4.6769755 |
| BI556771 | 76974 | 1190003J15Ri | 4.717042917 |
| BB559861 | 60510 | Syt9 | 4.853818083 |
| BB218107 | 16792 | Laptm5 | 4.885893 |
| NM_030676 | 26424 | Nr5a2 | 5.081557 |
| NM_008108 | 14562 | Gdf3 | 5.14758275 |
| NM_133982 | 102614 | Rpp25 | 5.190646483 |
| BG245669 | 14275 | Folr1 | 5.20975025 |
| NM_008652 | 17865 | Mybl2 | 5.210572917 |
| AV095095 | 17865 | Mybl2 | 5.318561367 |
| NM_009482 | 22286 | Utf1 | 5.421930317 |
| AK006110 | 67080 | 1700019D03R | 5.43684025 |
| C79957 | 13511 | Dsg2 | 5.656151683 |
| BB739342 | 58198 | Sall1 | 5.673027467 |
| BB414515 | 20527 | Slc2a3 | 5.699811883 |
| BG092030 | 13511 | Dsg2 | 6.010094817 |
| BG064756 | 99377 | Sall4 | 6.034042167 |
| BC006054 | 103551 | E130012A19R | 6.06322625 |
| AF220524 | 54427 | Dnmt3l | 6.105105033 |
| NM_009477 | 22271 | Upp1 | 6.128136833 |
| BB067210 | 73693 | Dppa4 | 6.187330683 |
| M75135 | 20527 | Slc2a3 | 6.2282474 |
| X69698 | 20527 | Slc2a3 | 6.23042505 |
| AV029604 | 22773 | Zic3 | 6.296894567 |
| NM_021480 | 58865 | Tdh | 6.529936483 |
| AF176530 | 50764 | Fbxo15 | 6.66547415 |
| AK005720 | 71827 | Lrrc34 | 6.691872117 |
| BB550860 | 21420 | Tcfap2c | 6.771666083 |
| AK010400 | 406217 | Bex4 | 6.816883167 |
| AV212609 | 353283 | Eras | 6.834029183 |
| AV298358 | 381269 | Mreg | 6.872264583 |
| NM_009337 | 21432 | Tcl1 | 6.879203683 |
| BB732077 | 22773 | Zic3 | 6.953696833 |
| AV099404 | 245128 | AU018091 | 7.054458833 |
| U31967 | 20674 | Sox2 | 7.296394917 |
| BB709552 | 14175 | Fgf4 | 7.45980965 |
| AV333667 | 26380 | Esrrb | 7.469081817 |
| NM_007430 | 11614 | Nr0b1 | 7.59146485 |
| BB041571 | 381591 | L1td1 | 7.663746183 |
| NM_009556 | 22702 | Zfp42 | 7.9192687 |
| NM_009864 | 12550 | Cdh1 | 7.9907572 |
| NM_013633 | 18999 | Pou5f1 | 8.090839917 |
| AK010332 | 71950 | Nanog | 8.503540967 |
| AV294613 | 21667 | Tdgf1 | 8.80162735 |

Table S4 Top 20 important miRNAs in stem cells

| |
|---|
| mmu-miR-290 |
| mmu-miR-293 |
| mmu-miR-291a-3p |
| mmu-miR-292-5p |
| mmu-miR-295 |
| mmu-miR-294 |
| mmu-miR-291a-5p |
| mmu-miR-291b-5p |
| mmu-miR-302b |
| mmu-miR-124a |
| mmu-miR-182 |
| mmu-miR-200a |
| mmu-miR-20b |
| mmu-miR-302d |
| mmu-miR-106a |
| mmu-miR-183 |
| mmu-miR-135b |
| mmu-miR-690 |
| mmu-miR-142-3p |
| mmu-miR-302 |